\documentclass[11pt,a4paper]{article}
\usepackage[english]{babel}
\usepackage{amsmath,amsthm,amssymb,epsfig,latexsym}
\usepackage{color}
\usepackage{ulem}
\usepackage[numbers,square,sort&compress]{natbib}

\newcommand{\so}{\scriptscriptstyle \rm I}
\newcommand{\st}{\scriptscriptstyle \rm I\hspace{-1pt}I}

\newcommand{\Phic}{\Phi^{\scriptscriptstyle C}}

\newcommand{\uc}{u^{\scriptscriptstyle C}}
\newcommand{\ub}{u^{\scriptscriptstyle B}}
\newcommand{\vc}{v^{\scriptscriptstyle C}}
\newcommand{\vb}{v^{\scriptscriptstyle B}}
\newcommand{\bu}{\bar u}
\newcommand{\bv}{\bar v}
\newcommand{\buc}{\bar{u}^{\scriptscriptstyle C}}
\newcommand{\bub}{\bar{u}^{\scriptscriptstyle B}}
\newcommand{\bvc}{\bar{v}^{\scriptscriptstyle C}}
\newcommand{\bvb}{\bar{v}^{\scriptscriptstyle B}}
\newcommand{\bucb}{\bar{u}^{\scriptscriptstyle C,B}}
\newcommand{\bvcb}{\bar{v}^{\scriptscriptstyle C,B}}

\newcommand{\bT}{\mathbb{T}}

\newcommand{\vac}{\Omega}
\newcommand{\dvac}{\Omega^\dagger}
\newcommand{\circN}{\overset{\;\circ}{\mathcal{N}}{}}

\newcommand{\be}[1]{\begin{equation}\label{#1}}
\newcommand{\ba}[1]{\begin{multline}\label{#1}}
\newcommand{\ee}{\end{equation}}
\newcommand{\ea}{\end{eqnarray}}

\newcommand{\diag}{\mathop{\rm diag}}

\newcommand{\str}{\mathop{\rm str}}

\newtheorem{prop}{Proposition}[section]

\newtheorem{Def}{Definition}[section]

\def\qed{\hfill\nobreak\hbox{$\square$}\par\medbreak}
 \makeatletter
 \@addtoreset{equation}{section}
 \makeatother
 
\newcommand{\bea}{\begin{eqnarray}}
\newcommand{\eea}{\end{eqnarray}}

\begin{document}
\renewcommand*{\thefootnote}{\fnsymbol{footnote}}
\begin{flushright}
LAPTH-035/16
\end{flushright}

\vspace{20pt}

\begin{center}
\begin{LARGE}
{\bf   Form factors of the monodromy matrix \\[1ex]
entries in $\mathfrak{gl}(2|1)$-invariant integrable models}
\end{LARGE}

\vspace{25pt}

\begin{large}
{A.~Hutsalyuk${}^{a}$,  A.~Liashyk${}^{b,c}$,
S.~Z.~Pakuliak${}^{a,d}$,\\ E.~Ragoucy${}^e$, N.~A.~Slavnov${}^f$\  \footnote{
hutsalyuk@gmail.com, a.liashyk@gmail.com, stanislav.pakuliak@jinr.ru, eric.ragoucy@lapth.cnrs.fr, nslavnov@mi.ras.ru}}
\end{large}

 \vspace{10mm}

${}^a$ {\it Moscow Institute of Physics and Technology,  Dolgoprudny, Moscow reg., Russia}

\vspace{3mm}

${}^b$ {\it Bogoliubov Institute for Theoretical Physics, NAS of Ukraine,  Kiev, Ukraine}

\vspace{3mm}

${}^c$ {\it National Research University Higher School of Economics,  Russia}

\vspace{3mm}

${}^d$ {\it Laboratory of Theoretical Physics, JINR,  Dubna, Moscow reg., Russia}

\vspace{3mm}

${}^e$ {\it Laboratoire de Physique Th\'eorique LAPTH, CNRS and Universit\'e de Savoie,\\
BP 110, 74941 Annecy-le-Vieux Cedex, France}

\vspace{3mm}

${}^f$ {\it Steklov Mathematical Institute of Russian Academy of Sciences, Moscow, Russia}

\end{center}

\vspace{2mm}


\vspace{1cm}

\begin{abstract}
We study  integrable models solvable by the nested algebraic Bethe ansatz and described by $\mathfrak{gl}(2|1)$  or $\mathfrak{gl}(1|2)$ superalgebras.
We obtain explicit determinant representations for form factors of the monodromy matrix entries. We show that all form factors are related to each other
at special limits of the Bethe parameters. Our results allow one to obtain determinant formulas for form factors of local operators in the supersymmetric
t-J model.
\end{abstract}

\vspace{2mm}


\renewcommand*{\thefootnote}{\arabic{footnote}}
\addtocounter{footnote}{-1}
\section{Introduction}

The algebraic Bethe ansatz is a powerful method of studying quantum integrable models \cite{FadST79,FadT79,FadLH96,KulR83}.
It can be used not only for finding spectra of quantum Hamiltonians, but for  an efficient  calculation of the form factors and correlation functions
as well \cite{IzeK84,BogIK93L,KitMT99,KitMST05,Sla07}.

The main objective of calculating the form factors of local operators in quantum integrable models is to provide
compact and manageable  representations for them. This problem was successfully solved in various integrable models
with $\mathfrak{gl}(2)$ symmetry and its $q$-deformation. There a determinant representation for the scalar product \cite{Sla89} of Bethe vectors was used
in order to obtain determinant formulas for form factors of the monodromy matrix entries. The latter directly leads to determinant representations
for form factors of local operators via the quantum inverse scattering problem \cite{KitMT99,MaiT00}. Furthermore, these determinants formulas allow
one to calculate the form factors of local operators even in the models for which the solution of the quantum inverse scattering problem is not known
\cite{Sla90,KorKS97}.
Determinant expressions for form factors were found to be very useful for analysis of correlation functions. They can be used either for
analytical calculations \cite{KitKMST11,KitKMST12} or for  numerical studies \cite{CauM05,CauPS07}.

Integrable models with higher rank symmetries were less studied, in spite of the generalization of the algebraic Bethe ansatz (nested
algebraic Bethe ansatz) was developed long ago  \cite{KulR83,Res86}. This was mainly due to technical difficulties in the study of such models. However,
recently determinant representations for form factors of local operators in the models with $\mathfrak{gl}(3)$-invariant $R$-matrix were obtained in the series of works \cite{PozOK12,BelPRS12b,BelPRS13a,PakRS14b,PakRS15a,PakRS15b}. Partial generalization of these results to the models with $q$-deformed
algebras was given in \cite{Sla15a}.

In this paper we study form factors of the monodromy matrix entries in the models described by $\mathfrak{gl}(2|1)$ and $\mathfrak{gl}(1|2)$
superalgebras. Actually, we focus mostly on the $\mathfrak{gl}(2|1)$ case, because the Yangians of these two superalgebras
$Y(\mathfrak{gl}(2|1))$ and $Y(\mathfrak{gl}(1|2))$ are related to each other by a simple isomorphism \cite{PakRS16a}.
In \cite{HutLPRS16c} we obtained a determinant representation for the scalar product of  special
(semi-on-shell) Bethe vectors in the models with $\mathfrak{gl}(2|1)$ symmetry.
There we also derived determinant formulas for form factors of diagonal entries of the monodromy matrix $T_{ii}$.
Using these results and the zero modes method \cite{PakRS15a} we obtain determinant representations for all form factors of the
operators $T_{ij}$. These formulas together with the inverse scattering problem \cite{GohK00} immediately lead to compact expressions for form
factors of local operators in the supersymmetric t-J model \cite{ZhaR88,And90,Sch87,For89,EssK92,FoeK93,Gom02}.

The article is organized as follows. In section~\ref{S-N} we introduce the model under consideration and describe
the notation used in the paper. We also define the form factors of the monodromy matrix entries  and describe some mappings  between them.
Section~\ref{S-DFFFF} contains the main results of the paper. Here we give determinant formulas for form factors of the monodromy matrix entries $T_{ij}$.
In section~\ref{S-PSS} we prove determinant representations for form factors of the diagonal entries $T_{ii}$ with respect to the same state.
In section~\ref{S-ZM} we introduce the zero modes of the operators $T_{ij}$ and derive
their action on Bethe vectors. Using these results we find additional relations between the different form factors. We show that all the form factors can be obtained from a single initial one by taking special limits of the Bethe parameters.
In section~\ref{S-detFF} we derive determinant representations for the  form factor of the off-diagonal monodromy matrix elements $T_{ij}$.
Finally, in section~\ref{S-FF-gl12} we apply our results to the models with $\mathfrak{gl}(1|2)$ symmetry.

\section{Notation and definitions\label{S-N}}

\subsection{Generalized $\mathfrak{gl}(2|1)$-invariant model}

The models considered below are described by the an
$R$-matrix acting in the tensor product $V_1\otimes V_2$ of two auxiliary spaces
$V_k\sim\mathbb{C}^{2|1}$, $k=1,2$ with the grading\footnote{Here and below we denote the grading in the $\mathfrak{gl}(2|1)$ superalgebra
by square brackets.} $[1]=[2]=0$, $[3]=1$.   Matrices acting in this space are also graded, according to $[e_{ij}]=[i]+[j]$, where $e_{ij}$ are elementary units: $(e_{ij})_{ab}=\delta_{ia}\delta_{jb}$.
The $R$-matrix has the following explicit form:
 \be{R-mat}
 R(x,y)=\mathbb{I}+g(x,y)P,\qquad g(x,y)=\frac{c}{x-y}.
 \ee
In the above definition, $\mathbb{I}$ is the identity matrix in $V_1\otimes V_2$, $P$ is the graded permutation matrix  \cite{KulS80}, and $c$ is a constant.

The monodromy matrix $T(w)$ satisfies the algebra
\be{RTT}
R(u,v)\bigl(T(u)\otimes \mathbb{I}\bigr) \bigl(\mathbb{I}\otimes T(v)\bigr)= \bigl(\mathbb{I}\otimes T(v)\bigr)
\bigl( T(u)\otimes \mathbb{I}\bigr)R(u,v).
\ee
Equation \eqref{RTT} holds in the graded tensor product $V_1\otimes V_2\otimes\mathcal{H}$,
where $\mathcal{H}$ is the Hilbert space of the Hamiltonian of the model under consideration.
The entries of the monodromy matrix $T(u)$ are graded in the same way as the matrices $e_{ij}$: $[T_{ij}(u)]=[i]+[j]$.
Being written in components, equation \eqref{RTT} takes the form
\be{RTT-comp}
\begin{aligned}
{}[T_{ij}(u),T_{kl}(v)\} & = (-1)^{[i]([k]+[l])+[k][l]}g(u,v)\Big(T_{kj}(v)\,T_{il}(u)-T_{kj}(u)\,T_{il}(v)\Big)\\
{}&= (-1)^{[l]([i]+[j])+[i][j]}g(u,v)\Big(T_{il}(u)\,T_{kj}(v)-T_{il}(v)\,T_{kj}(u)\Big),
\end{aligned}
\ee
where we have introduced a graded commutator
\be{Def-SupC}
[T_{ij}(u),T_{kl}(v)\}= T_{ij}(u)T_{kl}(v) -(-1)^{([i]+[j])([k]+[l])}   T_{kl}(v)  T_{ij}(u).
\ee

The supertrace in the auxiliary space $V\sim\mathbb{C}^{2|1}$ of the monodromy matrix,
\be{str}
\mathcal{T}(u)=\str T(u)=\sum_{i=1}^3 (-1)^{[i]}T_{ii}(u)
\ee
is called the transfer matrix. It is a generating
functional of the integrals of motion of the model. The eigenvectors of the transfer matrix are
called on-shell Bethe vectors (or simply on-shell vectors). They can be parameterized by sets of complex parameters
satisfying  Bethe equations (see section~\ref{SS-BV}).

Define a   linear mapping
\be{psi}
\psi\bigl(T_{ij}(u)\bigr)=(-1)^{[i][j]+[i]}T_{ji}(u),\qquad
\psi\bigl(AB\bigr)=(-1)^{[A][B]}\psi\bigl(B\bigr)\psi\bigl(A\bigr),
\ee
where $A$ and $B$ are arbitrary operators of fixed grading. The mapping \eqref{psi} is an
antimorphism of the algebra \eqref{RTT} \cite{PakRS16a}. It follows from \eqref{psi} that
\be{psiAn}
\psi\bigl(A_1\dots A_n\bigr)=(-1)^{\vartheta_n}\psi\bigl(A_n\bigr)\dots\psi\bigl(A_1\bigr), \qquad
\vartheta_n=\sum_{1\le i<j\le n} [A_i]\cdot [A_j].
\ee

\subsection{Notation}
We use the same notation and conventions as in the papers \cite{BelPRS13a,PakRS14b}.  We recall them for completeness.
Besides the function $g(x,y)$ we also introduce a function $f(x,y)$
\be{univ-not}
 f(x,y)=\frac{x-y+c}{x-y}.
\ee
Two other auxiliary functions  will be also used
\be{desand}
h(x,y)=\frac{f(x,y)}{g(x,y)}=\frac{x-y+c}{c},\qquad  t(x,y)=\frac{g(x,y)}{h(x,y)}=\frac{c^2}{(x-y)(x-y+c)}.
\ee
The following obvious properties of the functions introduced above are useful:
 \be{propert}
 \begin{aligned}
 &g(x,y)\sim \frac cx,\quad &h(x,y)\sim\frac xc,\quad  &f(x,y)\sim 1,\quad  &t(x,y)\sim\frac{c^2}{x^2},\quad &x \to\infty,\\
  &g(x,y)\sim -\frac cy,\quad &h(x,y)\sim-\frac yc,\quad  &f(x,y)\sim 1,\quad  &t(x,y)\sim\frac{c^2}{y^2},\quad &y \to\infty.
 \end{aligned}
 \ee

Before giving a description of the Bethe vectors we formulate a convention on the notations.
We denote sets of variables by bar: $\bar x$, $\bu$, $\bv$ etc.
Individual elements of the sets are denoted by  latin subscripts: $w_j$, $u_k$ etc.  The notation $\bu_i$,
means $\bu\setminus u_i$ etc. We say that $\bar x=\bar x'$,
if $\#\bar x=\#\bar x'$ and $x_i=x'_i$ (up to a permutation) for $i=1,\dots,\#\bar x$. We say that $\bar x\ne \bar x'$ otherwise.

In order to avoid too cumbersome formulas we use shorthand notation for products of  operators or
functions depending on one or two variables. Namely, if the functions $g$, $f$, and $h$  depend
on sets of variables, this means that one should take the product over the corresponding set.
For example,
 \be{SH-prod}
 h(\bar u,v)=\prod_{u_j\in\bar u} h(u_j,v);\quad
  g(z, \bar x_i)= \prod_{\substack{x_j\in\bar x\\x_j\ne x_i}} g(z, x_j);\quad
 f(\bu,\bv)=\prod_{u_j\in\bu}\prod_{v_k\in\bv} f(u_j,v_k).
 \ee
This notation is also used for the product of commuting operators,
\be{SH-prod-O}
T_{ij}(\bar u)=\prod_{u_k\in\bar u} T_{ij}(u_k), \qquad\text{if}\quad [i]+[j]=0,\quad \mod(2).
\ee
One can easily see from the commutation relations \eqref{RTT-comp} that in this case $[T_{ij}(u),T_{ij}(v)]=0$, and hence, the
operator product \eqref{SH-prod-O} is well defined. However, if $[i]+[j]=1$, then $[T_{ij}(u),T_{ij}(v)]\ne 0$, therefore we
introduce symmetric operator products
\be{bTc-def}
\bT_{j3}(\bv)= \frac{T_{j3}(v_1)\dots T_{j3}(v_n)}{\prod_{n\ge \ell>m\ge 1} h(v_\ell,v_m)},
\qquad \bT_{3j}(\bv)= \frac{T_{3j}(v_1)\dots T_{3j}(v_n)}{\prod_{n\ge \ell>m\ge 1} h(v_m,v_\ell)} \qquad j=1,2.
\ee
It is easy to check that if $[i]=[j]=0$, then
\be{psi-bT1}
\psi\bigl(T_{ij}(\bu)\bigr)= T_{ji}(\bu),\quad
\psi\bigl(\bT_{i3}(\bu)\bigr)=(-1)^{n(n-1)/2} \bT_{3i}(\bu), \quad
\psi\bigl(\bT_{3i}(\bu)\bigr)=(-1)^{n(n+1)/2} \bT_{i3}(\bu),
\ee
where $n=\#\bu$.

\subsection{Bethe vectors\label{SS-BV}}

Now we pass to the description of Bethe vectors.
A generic Bethe vector is denoted by $\mathbb{B}_{a,b}(\bu;\bv)$.
It is parameterized by two sets of
complex parameters $\bu=u_1,\dots,u_a$ and $\bv=v_1,\dots,v_b$ with $a,b=0,1,\dots$. They are called Bethe parameters.
Dual Bethe vectors are denoted by $\mathbb{C}_{a,b}(\bu;\bv)$. They also depend on two sets of
complex parameters $\bu=u_1,\dots,u_a$ and $\bv=v_1,\dots,v_b$. The state with
$\bu=\bv=\emptyset$ is called a pseudovacuum vector $\vac$. Similarly, the dual state
with $\bu=\bv=\emptyset$ is called a dual pseudovacuum vector $\dvac$. These vectors
are annihilated by the operators $T_{ij}(w)$, where $i>j$ for  $\vac$ and $i<j$ for $\dvac$.
At the same time both vectors are eigenvectors for the diagonal entries of the monodromy matrix
 \be{Tjj}
 T_{ii}(w)\vac=\lambda_i(w)\vac, \qquad   \dvac T_{ii}(w)=\lambda_i(w)\dvac,\quad i=1,2,3,
 \ee
where $\lambda_i(w)$ are some scalar functions. In the framework of the generalized model, $\lambda_i(w)$ remain free functional parameters.
Below we will often deal with ratios
 \be{ratios}
 r_1(w)=\frac{\lambda_1(w)}{\lambda_2(w)}, \qquad  r_3(w)=\frac{\lambda_3(w)}{\lambda_2(w)}.
 \ee
We extend the convention on the shorthand notation for the products to the functions $\lambda_i(w)$ and $r_k(w)$, for instance,
\be{SH-prodlr}
r_1(\bu)=\prod_{u_i\in \bu}r_1(u_i),\qquad \lambda_2(\bv_j)=\prod_{\substack{v_i\in \bv\\ v_i\ne v_j}}\lambda_2(v_i).
\ee

Bethe vectors in the models described by superalgebras were studied in \cite{BelR08}.
There exist several explicit formulas for the Bethe vectors in terms of polynomials in $T_{ij}(w)$ (with $i<j$) acting
on the pseudovacuum $\vac$ (see \cite{PakRS16a}). We give here one of those representations in order to fix
normalization.
For $\#\bu=a$ and $\#\bv=b$ we define a  Bethe vector  $\mathbb{B}_{a,b}(\bu;\bv)$ and its dual vector $\mathbb{C}_{a,b}(\bu;\bv)$ as
\be{BV-expl1}
\mathbb{B}_{a,b}(\bu;\bv)=\sum \frac{g(\bv_{\so},\bu_{\so})
f(\bu_{\so},\bu_{\st}) g(\bv_{\st},\bv_{\so})h(\bu_{\so},\bu_{\so})}{f(\bv,\bu)\lambda_2(\bu)\lambda_2(\bv_{\st})} \;
\bT_{13}(\bu_{\so})\,T_{12}(\bu_{\st})\,\bT_{23}(\bv_{\st})\vac,
\ee
\be{dBV-expl1}
\mathbb{C}_{a,b}(\bu;\bv)=(-1)^{b(b-1)/2}\sum \frac{g(\bv_{\so},\bu_{\so})
f(\bu_{\so},\bu_{\st}) g(\bv_{\st},\bv_{\so})h(\bu_{\so},\bu_{\so})}{f(\bv,\bu) \lambda_2(\bu)\lambda_2(\bv_{\st}) }\;\dvac
\bT_{32}(\bv_{\st})\,T_{21}(\bu_{\st})\,\bT_{31}(\bu_{\so}).
\ee
Here the sum is taken over partitions of the set $\bv$ into two disjoint subsets $\bv_{\so}$ and $\bv_{\st}$ and over partitions of $\bu$ into  disjoint subsets $\bu_{\so}$ and $\bu_{\st}$. The partitions are independent except that  $\#\bu_{\so}=\#\bv_{\so}=n$, where $n=0,1,\dots,\min(a,b)$. Recall also that we use the shorthand notation for the products of all the functions and
the operators in \eqref{BV-expl1}, \eqref{dBV-expl1}. Observe, that (dual) Bethe vectors are symmetric over $\bu$ and symmetric over $\bv$.

If the parameters $\bu$ and $\bv$ of a Bethe vector\footnote{%
For simplicity here and below we do not distinguish between vectors and dual vectors.}
satisfy a special system of equations (Bethe equations), then
it becomes an eigenvector of the transfer matrix (on-shell Bethe vector). The system of Bethe equations can be written in the following form:
\be{AEigenS-1}
\begin{aligned}
r_1(u_{i})&=\frac{f(u_{i},\bu_{i})}{f(\bu_{i},u_{i})}f(\bv,u_{i}),\qquad i=1,\dots,a,\\
r_3(v_{j})&=f(v_{j},\bu), \qquad j=1,\dots,b.
\end{aligned}
\ee
Recall that $\bu_{i}=\bu\setminus u_i$ and $\bv_j=\bv\setminus v_j$.

If $\bu$ and $\bv$ satisfy the system \eqref{AEigenS-1}, then
\be{Left-act}
\mathcal{T}(z)\mathbb{B}_{a,b}(\bu;\bv) = \tau(z|\bu,\bv)\,\mathbb{B}_{a,b}(\bu;\bv),\qquad
\mathbb{C}_{a,b}(\bu;\bv)\mathcal{T}(z) = \tau(z|\bu,\bv)\,\mathbb{C}_{a,b}(\bu;\bv),
\ee
where $\mathcal{T}(z)$ is defined by \eqref{str} and
\be{tau-def}
\tau(z)\equiv\tau(z|\bu,\bv)=\lambda_1(z)f(\bu,z)+\lambda_2(z)f(z,\bu)f(\bv,z)-\lambda_3(z)f(\bv,z).
\ee

 {\sl Remark.\label{GenMod1}} In concrete quantum models the functions $r_1(z)$ and $r_3(z)$ are fixed. Then the system of Bethe equations
\eqref{AEigenS-1} determines the admissible values of the parameters $\bu$ and $\bv$. Eventually these values characterize the spectrum
of the Hamiltonian of the quantum model under consideration.  However, in the generalized model, where  $r_1(z)$ and $r_3(z)$ are free functional
parameters, the situation is opposite. The system \eqref{AEigenS-1} only fixes the values of the functions $r_1(z)$ and $r_3(z)$ in several
points, while the parameters $\bu$ and $\bv$ remain arbitrary complex numbers \cite{Kor82}.

Apart from the usual transfer matrix it is often convenient to consider a twisted transfer matrix \cite{IzeK84,KitMST05,BelPRS12b,HutLPRS16c}
$\mathcal{T}_\kappa(z)=\str\bigl(\hat\kappa T(z)\bigr)$, where $\hat\kappa=\diag(\kappa_1,\kappa_2,\kappa_3)$ and $\kappa_i$ are some complex numbers. Its eigenvectors
\be{K-Left-act}
\mathcal{T}_\kappa(z)\mathbb{B}^{(\kappa)}_{a,b}(\bu;\bv) = \tau_\kappa(z|\bu,\bv)\,\mathbb{B}^{(\kappa)}_{a,b}(\bu;\bv),\qquad
\mathbb{C}^{(\kappa)}_{a,b}(\bu;\bv)\mathcal{T}_\kappa(z) = \tau_\kappa(z|\bu,\bv)\,\mathbb{C}^{(\kappa)}_{a,b}(\bu;\bv),
\ee
are called twisted on-shell Bethe vectors. The parameters of  these vectors satisfy a system of twisted Bethe equations
\be{K-AEigenS-1}
\begin{aligned}
r_1(u_{j})&=\frac{\kappa_2}{\kappa_1}\frac{f(u_{j},\bu_{j})}{f(\bu_{j},u_{j})}f(\bv,u_{j}),\qquad j=1,\dots,a,\\
r_3(v_{j})&=\frac{\kappa_2}{\kappa_3}f(v_{j},\bu), \qquad j=1,\dots,b.
\end{aligned}
\ee
The twisted eigenvalue $\tau_\kappa(z)$ has the form
\be{K-tau-def}
\tau_\kappa(z)\equiv\tau_\kappa(z|\bu,\bv)= \kappa_1\lambda_1(z)f(\bu,z)+\kappa_2\lambda_2(z)f(z,\bu)f(\bv,z)-\kappa_3\lambda_3(z)f(\bv,z).
\ee

The norm of the Bethe vector defined above was calculated in \cite{HutLPRS16c}
\be{Norm-res}
\mathbb{C}_{a,b}(\bu;\bv)\mathbb{B}_{a,b}(\bu;\bv)=(-c)^{a+b}\prod_{j=1}^b\prod_{k=1}^a f(v_j,u_k)\prod_{\substack{j,k=1\\j\ne k}}^a f(u_j,u_k)\;
\prod_{\substack{j,k=1\\j\ne k}}^b g(v_j,v_k)\;\det_{a+b}\widehat{\mathcal{N}}.
\ee
The determinant in \eqref{Norm-res} is nothing but the Jacobian of the Bethe equations \eqref{AEigenS-1} in the logarithmic form
(see also \cite{Kor82,BelPRS12b}). Let
\be{Phi}
\begin{aligned}
&\Phi_j= \log\Big(\frac{r_1(u_j)}{f(\bv,u_j)}\frac{f(\bu_j,u_j)}{f(u_j,\bu_j)} \Big),\qquad j=1,\dots,a,\\
&\Phi_{a+j}=\log\left(\frac{r_3(v_j)}{f(v_j,\bu)}\right),\qquad j=1,\dots,b.
\end{aligned}
\ee
Then
\be{Norm-Phi}
\widehat{\mathcal{N}}_{j,k}=\frac{\partial\Phi_j}{\partial x_k}
\qquad j,k=1,\dots,a+b,
\ee
and $\{x_1,\dots,x_{a+b}\}=\{u_1,\dots,u_a,v_1,\dots,v_b\}$.

It is convenient to extend the action of the antimorphism $\psi$ on vectors.  We can always choose the
grading of $\vac$ and $\dvac$ such that
$[\vac]=[\dvac]=0$. Then we set $[A\vac]=[\dvac A]=[A]$ and define
\be{act-vec}
\begin{aligned}
&\psi(\vac)=\dvac,\qquad  \psi(A\vac)=\dvac\psi(A),\\
&\psi(\dvac)=\vac,\qquad  \psi(\dvac A)=\psi(A)\vac,
\end{aligned}
\ee
where $A$ is an arbitrary product of the monodromy matrix entries.
It is easy to see that
\be{par-BV}
[\mathbb{B}_{a,b}(\bu;\bv)]=[\mathbb{C}_{a,b}(\bu;\bv)]=b.
\ee
One can also convince oneself that
\be{psi-BvdBV}
\psi\bigl(\mathbb{B}_{a,b}(\bu;\bv)\bigr)=\mathbb{C}_{a,b}(\bu;\bv),\qquad
\psi\bigl(\mathbb{C}_{a,b}(\bu;\bv)\bigr)=(-1)^b\,\mathbb{B}_{a,b}(\bu;\bv).
\ee
Indeed, let us fix the partitions in \eqref{BV-expl1}, \eqref{dBV-expl1} such that $\#\bu_{\so}=\#\bv_{\so}=n$. Then using
\eqref{psi-bT1} we find
\begin{multline}\label{psi-BV0}
\psi\Bigl(\bT_{13}(\bu_{\so})\,T_{12}(\bu_{\st})\,\bT_{23}(\bv_{\st})\Bigr)=(-1)^{n(n-1)/2+(b-n)(b-n-1)/2+n(b-n)}
\bT_{32}(\bv_{\st})\,T_{21}(\bu_{\st})\,\bT_{31}(\bu_{\so})\\
=(-1)^{b(b-1)/2}
\bT_{32}(\bv_{\st})\,T_{21}(\bu_{\st})\,\bT_{31}(\bu_{\so}),
\end{multline}
and similarly
\begin{multline}\label{psi-dBV0}
\psi\Bigl(\bT_{32}(\bv_{\st})\,T_{21}(\bu_{\st})\,\bT_{31}(\bu_{\so})\Bigr)=(-1)^{n(n+1)/2+(b-n)(b-n+1)/2+n(b-n)}
\bT_{13}(\bu_{\so})\,T_{12}(\bu_{\st})\,\bT_{23}(\bv_{\st})\\
=(-1)^{b(b+1)/2}\bT_{13}(\bu_{\so})\,T_{12}(\bu_{\st})\,\bT_{23}(\bv_{\st}).
\end{multline}
These equations immediately imply \eqref{psi-BvdBV}.

\subsection{Form factors of the monodromy matrix entries\label{SS-FF}}

We define form factors of the monodromy matrix entries as
\be{FF-def0}
\mathcal{F}^{(i,j)}\left(z\Bigl|\begin{smallmatrix}
\buc & \bub \\ \bvc & \bvb \end{smallmatrix}\right)^{a',a}_{b',b}=\mathbb{C}_{a',b'}(\buc;\bvc)T_{ij}(z)\mathbb{B}_{a,b}(\bub;\bvb).
\ee
Here both $\mathbb{C}_{a',b'}(\buc;\bvc)$ and $\mathbb{B}_{a,b}(\bub;\bvb)$ are on-shell
Bethe vectors, the parameter $z$ is an arbitrary complex  number, and
\be{apabpb}
\begin{array}{l}
a'=a+\delta_{i1}-\delta_{j1},\\
b'=b+\delta_{j3}-\delta_{i3}.
\end{array}
\ee
Similarly to the $\mathfrak{gl}(3)$ case, one can also introduce universal form factors \cite{PakRS15a}, if $\{\buc,\bvc\}\ne \{\bub,\bvb\}$.
Namely, let
\be{UFF-def0}
\mathfrak{F}^{(i,j)}\left(\begin{smallmatrix}
\buc & \bub \\ \bvc & \bvb
\end{smallmatrix}\right)^{a',a}_{b',b}
=\frac{\mathcal{F}^{(i,j)}\left(z\Bigl|\begin{smallmatrix}
\buc & \bub \\ \bvc & \bvb\end{smallmatrix}\right)^{a',a}_{b',b}}
{\tau(z|\buc,\bvc)-\tau(z|\bub,\bvb)}.
\ee
It is easy to show that the functions $\mathfrak{F}^{(ij)}$ do not depend on $z$. Indeed, it follows from the
commutation relations \eqref{RTT-comp} that
\be{strT1}
[\mathcal{T}(z) , T_{ij}(w)]=
[\mathcal{T}(w) , T_{ij}(z)],
\ee
where $\mathcal{T}$ is the transfer matrix \eqref{str}.
Hence, for arbitrary on-shell Bethe vectors $\mathbb{C}_{a',b'}(\buc;\bvc)$ and $\mathbb{B}_{a,b}(\bub;\bvb)$ we obtain
\be{strT2}
\mathbb{C}_{a',b'}(\buc;\bvc)[\mathcal{T}(z) , T_{ij}(w)]\mathbb{B}_{a,b}(\bub;\bvb)=
\mathbb{C}_{a',b'}(\buc;\bvc)[\mathcal{T}(w) , T_{ij}(z)]\mathbb{B}_{a,b}(\bub;\bvb).
\ee
Using \eqref{Left-act} we find
\begin{multline}\label{strT3}
\bigl(\tau(z|\buc,\bvc)-\tau(z|\bub,\bvb)\bigr) \mathbb{C}_{a',b'}(\buc;\bvc)  T_{ij}(w)\mathbb{B}_{a,b}(\bub;\bvb)\\
=\bigl(\tau(w|\buc,\bvc)-\tau(w|\bub,\bvb)\bigr) \mathbb{C}_{a',b'}(\buc;\bvc)  T_{ij}(z)\mathbb{B}_{a,b}(\bub;\bvb),
\end{multline}
where $\tau$ are eigenvalues of the transfer matrix. Equation \eqref{strT3} immediately yields
\be{strT4}
\frac{ \mathbb{C}_{a',b'}(\buc;\bvc)  T_{ij}(w)\mathbb{B}_{a,b}(\bub;\bvb)}{\tau(w|\buc,\bvc)-\tau(w|\bub,\bvb)}
=\frac{ \mathbb{C}_{a',b'}(\buc;\bvc)  T_{ij}(z)\mathbb{B}_{a,b}(\bub;\bvb)}
{\tau(z|\buc,\bvc)-\tau(z|\bub,\bvb)}.
\ee
We see that the l.h.s. of \eqref{strT4} depends on $w$, while the r.h.s.  depends on $z$. Thus, the ratio \eqref{UFF-def0}
does not depend on the argument of the operator $T_{ij}$.

We call the form factors \eqref{UFF-def0} universal, because they are determined by the $R$-matrix only. In other words, for a given
$R$-matrix they do not depend on the monodromy matrix, and hence, they are model independent. Indeed, all the dependence of the
form factors on a specific model is hidden in the functions $r_1$ and $r_3$.
More specifically, since the dependance on $r_1(u_i)$ and $r_3(v_i)$ can be removed using the Bethe equations \eqref{AEigenS-1}, the real dependance in the model is concentrated in the terms $r_1(z)$ and $r_3(z)$. Since the universal form factors do not depend on $z$, they cannot depend on $r_1(z)$ and $r_3(z)$.
Thus, as we have claimed above, they do not depend on
the monodromy matrix of the model.

{\sl Remark.} Strictly speaking the universal form factors do not depend on the functions $r_k$, if $\buc\cap\bub=\emptyset$ and
$\bvc\cap\bvb=\emptyset$, that is when the Bethe parameters of both vectors are  all
different. Otherwise,  if, for instance, $\uc_j=\ub_k$,
then the universal form factors depend on the logarithmic derivative $\log'r_1(\ub_k)$ of the function $r_1(u)$ \cite{PakRS15a}. Similarly,
if $\vc_j=\vb_k$,
then the universal form factors depend on the logarithmic derivative $\log'r_3(\vb_k)$ of the function $r_3(v)$.

\begin{prop}\label{PFF-FF}
Form factors $\mathcal{F}^{(i,j)}$ and $\mathcal{F}^{(j,i)}$ are related by
\be{FF-FFprop}
\mathcal{F}^{(i,j)}\left(z\Bigl|\begin{smallmatrix}
\buc & \bub \\ \bvc & \bvb \end{smallmatrix}\right)^{a',a}_{b',b}=(-1)^{\theta_{ij}}
\mathcal{F}^{(j,i)}\left(z\Bigl|\begin{smallmatrix}
\bub & \buc \\ \bvb & \bvc \end{smallmatrix}\right)^{a,a'}_{b,b'},
\ee
where
\be{Theta}
\begin{aligned}
&\theta_{ij}=0,\qquad &[i]+[j]=0,\qquad&\mod(2),\\
&\theta_{ij}=b,\qquad &[i]=0,\quad [j]=1,&{}\\
&\theta_{ij}=b+1,\qquad &[i]=1,\quad [j]=0.&{}
\end{aligned}
\ee
\end{prop}

{\sl Proof}. Since a form factor $\mathcal{F}^{(i,j)}$ is a $c$-number function, it is invariant under the action of
the antimorphism $\psi$:
\be{psi-FFl}
\psi\left( \mathcal{F}^{(i,j)}\left(z\Bigl|\begin{smallmatrix}
\buc & \bub \\  \bvc & \bvb     \end{smallmatrix}  \right)^{a',a}_{b',b}\right)=
\mathcal{F}^{(i,j)}\left(z\Bigl| \begin{smallmatrix}
\buc &\bub  \\ \bvc & \bvb \end{smallmatrix}  \right)^{a',a}_{b',b}.
\ee
On the other hand, acting with $\psi$ on the r.h.s. of \eqref{FF-def0} we obtain
\begin{multline}\label{psi-FFr}
\psi\Bigl(  \mathbb{C}_{a',b'}(\buc;\bvc)T_{ij}(z)\mathbb{B}_{a,b}(\bub;\bvb)\Bigr)\\
=(-1)^{([i]+[j])(b+b')+b'b}
\psi\Bigl( \mathbb{B}_{a,b}(\bub;\bvb)\Bigr)\psi\Bigl(T_{ij}(z) \Bigr)\psi\Bigl(  \mathbb{C}_{a',b'}(\buc;\bvc) \Bigr)\\
=(-1)^{\theta_{ij}}\mathbb{C}_{a,b}(\bub;\bvb)T_{ji}(z) \mathbb{B}_{a',b'}(\buc;\bvc),
\end{multline}
where we used \eqref{psi}, \eqref{par-BV},  \eqref{psi-BvdBV}, and
\be{th-0}
\theta_{ij}=([i]+[j])(b+b')+b'b+b'+[i][j]+[i].
\ee
Thus, we have reduced the form factor $\mathcal{F}^{(i,j)}$ to the form factor $\mathcal{F}^{(j,i)}$. In order to simplify the
phase factor we can use \eqref{apabpb}
\be{apabpb0}
b'-b=\delta_{j3}-\delta_{i3}=[j]-[i].
\ee
After elementary algebra we obtain
\be{th-1}
\theta_{ij}=([j]+[i])b+[i][j]+[i],\qquad \mod(2),
\ee
and it is straightforward to check that this expression is equivalent to \eqref{Theta}. \qed

It follows from \eqref{FF-FFprop} that form factors of diagonal matrix elements $\mathcal{F}^{(i,i)}$ are invariant under
the replacement $\buc\leftrightarrow\bub$ and $\bvc\leftrightarrow\bvb$. This invariance yields the following transformation
of the corresponding universal form factors
\begin{equation}\label{Fii-ii}
\mathfrak{F}^{(i,i)}\left(\begin{smallmatrix}
\buc & \bub \\ \bvc & \bvb \end{smallmatrix}\right)^{a,a}_{b,b}=
-\mathfrak{F}^{(i,i)}\left(\begin{smallmatrix}
\bub & \buc \\ \bvb & \bvc \end{smallmatrix}\right)^{a,a}_{b,b}.
\end{equation}
The  minus sign appears due to the  denominator in \eqref{UFF-def0}. For the universal form factors
of the off-diagonal matrix elements we obtain
\begin{equation}\label{F31-13}
\mathfrak{F}^{(3,1)}\left(\begin{smallmatrix}
\buc & \bub \\ \bvc & \bvb \end{smallmatrix}\right)^{a,a+1}_{b,b+1}=(-1)^{b+1}
\mathfrak{F}^{(1,3)}\left(\begin{smallmatrix}
\bub & \buc \\ \bvb & \bvc \end{smallmatrix}\right)^{a+1,a}_{b+1,b},
\end{equation}
\begin{equation}\label{F32-23}
\mathfrak{F}^{(3,2)}\left(\begin{smallmatrix}
\buc & \bub \\ \bvc & \bvb \end{smallmatrix}\right)^{a,a}_{b,b+1}=(-1)^{b+1}
\mathfrak{F}^{(2,3)}\left(\begin{smallmatrix}
\bub & \buc \\ \bvb & \bvc \end{smallmatrix}\right)^{a,a}_{b+1,b},
\end{equation}
and
\begin{equation}\label{F21-12}
\mathfrak{F}^{(2,1)}\left(\begin{smallmatrix}
\buc & \bub \\ \bvc & \bvb \end{smallmatrix}\right)^{a,a+1}_{b,b}=
-\mathfrak{F}^{(1,2)}\left(\begin{smallmatrix}
\bub & \buc \\ \bvb & \bvc \end{smallmatrix}\right)^{a+1,a}_{b,b}.
\end{equation}

\section{Determinant formulas for form factors\label{S-DFFFF}}

Considering form factors of the monodromy matrix entries one should distinguish between two cases\footnote{%
Here and below for brevity we write $\{\buc,\bvc\}=\{\bub,\bvb\}$, although one should understand this condition as
$\buc=\bub$  and $\bvc=\bvb$.}: (1)
$\{\buc,\bvc\}=\{\bub,\bvb\}$; (2) $\{\buc,\bvc\}\ne\{\bub,\bvb\}$. The first case occurs  only for form factors
$\mathcal{F}^{(i,i)}$ of diagonal matrix elements $T_{ii}(z)$. Indeed, the condition $\{\buc,\bvc\}=\{\bub,\bvb\}$
implies  $a'=a$ and $b'=b$ (see \eqref{apabpb}), which is possible  for diagonal entries $T_{ii}(z)$ only. We first present
the results for this case.

\subsection{Form factors between  identical  states\label{SS-FFSS}}

Let $\buc=\bub=\bu$ and $\bvc=\bvb=\bv$.
The form factors $\mathcal{F}^{(i,i)}$ have the following determinant representations:
\be{FF-res1}
\mathcal{F}^{(i,i)}\left(z\Bigl|\begin{smallmatrix}
\bu & \bu \\ \bv & \bv \end{smallmatrix}\right)^{a,a}_{b,b}=(-c)^{a+b}\prod_{j=1}^b\prod_{k=1}^a f(v_j,u_k)\prod_{\substack{j,k=1\\j\ne k}}^a f(u_j,u_k)\;
\prod_{\substack{j,k=1\\j\ne k}}^b g(v_j,v_k)\;\det_{a+b+1}\widehat{\mathcal{N}}^{(i,i)}.
\ee
In order to describe the $(a+b+1)\times (a+b+1)$ matrices $\widehat{\mathcal{N}}^{(i,i)}$ we combine the sets $\bu$ and $\bv$ into a set
$\bar x=\{u_1,\dots,u_a,v_1,\dots,v_b\}$. Then
%
%
%
%
%
\be{hatNii}
\begin{aligned}
&\widehat{\mathcal{N}}^{(i,i)}_{j,k}= \frac{\partial\Phi_j}{\partial x_k}
\qquad &j,k=1,\dots,a+b,\\
&\widehat{\mathcal{N}}^{(i,i)}_{a+b+1,k}= (-1)^{[i]}\frac{\partial\tau(z|\bu,\bv)}{\partial x_k},\qquad & k=1,\dots,a+b,\\
&\widehat{\mathcal{N}}^{(i,i)}_{j,a+b+1}=\delta_{i1}-\delta_{i2},\qquad & j=1,\dots,a,\\
&\widehat{\mathcal{N}}^{(i,i)}_{j,a+b+1}=\delta_{i3}-\delta_{i2},\qquad & j=a+1,\dots,a+b,\\
&\widehat{\mathcal{N}}^{(i,i)}_{a+b+1,a+b+1}= (-1)^{[i]}\frac{\partial\tau_\kappa(z|\bu,\bv)}{\partial \kappa_i}.\qquad & {}
\end{aligned}
\ee
Here $\Phi_j$ are given by \eqref{Phi}, and the eigenvalues of the usual and twisted transfer matrices $\tau(z|\bu,\bv)$ and
$\tau_\kappa(z|\bu,\bv)$ are defined respectively in \eqref{tau-def} and \eqref{K-tau-def}.
The proof of the determinant formula \eqref{FF-res1} will be given in section~\ref{S-PSS}.

\subsection{Form factors between different states\label{SS-FFDS}}

\subsubsection{Notation}

If $\{\buc,\bvc\}\ne\{\bub,\bvb\}$, then the universal form factors are well defined. We assume that the sets of Bethe parameters $\buc$, $\bvc$, $\bub$, $\bvb$ are fixed and their cardinalities are
\be{Card-BP}
\#\buc=a',\qquad\#\bub=a,\qquad\#\bvc=b',\qquad\#\bvb=b,
\ee
where $a'$ and $b'$ are related to $a$ and $b$ by \eqref{apabpb}.
Before giving explicit determinant presentations for the universal form factors we introduce several new functions.

For a set of variables $\bar w=\{w_1,\dots,w_n\}$ define
\be{def-Del}
\Delta'(\bar w)
=\prod_{j<k}^n g(w_j,w_k),\qquad {\Delta}(\bar w)=\prod_{j>k}^n g(w_j,w_k).
\ee
Then we introduce a function
\begin{equation}\label{Hab0}
H(\buc;\bub;\bvc)= f(\bvc,\bub)h(\bub,\bub)\Delta'(\buc)\Delta(\bub)\Delta(\bvc)\Delta'(\bvc).
\end{equation}
The function $H$ plays the role of a universal prefactor that appears in all determinant formulas for form factors.
One should remember, however, that in spite of this function has the universal representation \eqref{Hab0}, the cardinalities
of the sets $\buc$, $\bub$, and $\bvc$ are different for the different form factors.

We define also a $(a'+b')$-component vector $\Omega$ as
\be{vector}
\begin{aligned}
\Omega_j&=\frac{g(\uc_j,\buc_j)}{g(\uc_j,\bub)},\qquad j=1,\dots,a',\\
\Omega_{a'+j}&=\frac{g(\vc_j,\bvc_j)}{g(\vc_j,\bvb)},\qquad j=1,\dots,b'.
\end{aligned}
\ee
Since we consider the case $\{\buc,\bvc\}\ne\{\bub,\bvb\}$,  there exists at least one  component
$\Omega_p$ such that  $\Omega_p\ne 0$.

Finally, for fixed sets of variables $\buc$, $\bub$, $\bvc$, and $\bvb$
we  introduce two  rectangular matrices $\mathcal{L}$ and $\mathcal{M}$.
The matrix $\mathcal{L}$ has the size $a'\times(a+b')$ and its entries are
\begin{equation}\label{matMup}
 \mathcal{L}_{j,k} = t(\uc_j,x_k)\frac{(-1)^{a'-1}r_1(x_k)h(\buc,x_k)}{f(\bvc,x_k)h(x_k,\bub)} + t(x_k,\uc_j) \frac{h(x_k,\buc)}{h(x_k,\bub)},
 \qquad \begin{array}{l}j=1,\dots,a', \\
  k=1,\dots, a+b'.\end{array}
\end{equation}
The matrix $\mathcal{M}$ has the size $b'\times(a+b')$ and its entries are
\begin{equation}\label{matMdown}
 \mathcal{M}_{j,k} = - t(\vc_j,x_k) \frac{g(\bvb,x_k)}{g(\bvc,x_k)}\left(1-\frac{r_3(x_k)}{f(x_k,\bub)}\right), \qquad
\begin{array}{l}j=1,\dots,b', \\
  k=1,\dots,a+b'.\end{array}
\end{equation}
Here the set $\bar x$ is the union of two sets: $\bar x=\{\bub,\bvc\}$.
Actually, both matrices $\mathcal{L}$ and $\mathcal{M}$ consist of two blocks depending on whether $x_k\in\bub$
or $x_k\in\bvc$. The structures of these blocks are very different,  and we give now a more detailed description of
them.

First of all, we note that $1/f(\bvc,x_k)=0$ if $x_k\in\bvc$, and $1/f(x_k,\bub)=0$ if $x_k\in\bub$. Therefore we obtain
\begin{equation}\label{L-anti}
 \mathcal{L}_{j,k+a} =  t(\vc_k,\uc_j) \frac{h(\vc_k,\buc)}{h(\vc_k,\bub)},
 \qquad   k=1,\dots, b',
\end{equation}
and
\begin{equation}\label{M-anti}
 \mathcal{M}_{j,k} = - t(\vc_j,\ub_k) \frac{g(\bvb,\ub_k)}{g(\bvc,\ub_k)}, \qquad
  k=1,\dots,a.
\end{equation}

The product $1/g(\bvc,x_k)$ also vanishes, if $x_k\in\bvc$. However, this zero can be compensated by the pole of the function
$t(\vc_j,x_k)$, if $x_k=\vc_j$. Therefore, the block of the matrix $\mathcal{M}$ with $k>a$ has diagonal structure:
\begin{equation}\label{M-diag}
 \mathcal{M}_{j,a+k} = - \delta_{jk} \frac{g(\bvb,\vc_k)}{g(\bvc_k,\vc_k)}\left(1-\frac{f(\vc_k,\buc)}{f(\vc_k,\bub)}\right), \qquad
  k=1,\dots,b'.
\end{equation}
Here we replaced the function $r_3(\vc_k)$ with the product $f(\vc_k,\buc)$ due to the Bethe equations. One should remember, however, that
this replacement is possible only if $\bvc\cap\bvb=\emptyset$. Otherwise, if some parameters $\vc_{j_1},\dots,\vc_{j_\ell}$ from the set $\bvc$ coincide
with the parameters $\vb_{j_1},\dots,\vb_{j_\ell}$
from the set $\bvb$, then one should first take the limits $\vc_{j_s}\to\vb_{j_s}$ in \eqref{matMdown} and only after this we can impose Bethe
equations for the functions $r_3(\vc_k)$.

Similarly, if $\buc\cap\bub=\emptyset$, then the matrix elements $ \mathcal{L}_{j,k}$ with $j=1,\dots,a'$ and $k=1,\dots, a$
take the form
\begin{equation}\label{L-diag}
 \mathcal{L}_{j,k} = (-1)^{a'+a}t(\uc_j,\ub_k)\frac{f(\bvb,\ub_k)h(\buc,\ub_k)}{f(\bvc,\ub_k)h(\bub,\ub_k)} + t(\ub_k,\uc_j) \frac{h(\ub_k,\buc)}{h(\ub_k,\bub)}.
\end{equation}

\subsubsection{Determinant formulas}

Now we give the list of determinant representations for the universal form factors of the matrix elements
$T_{ij}(z)$. Certainly, it should be enough to give explicit formulas for $\mathfrak{F}^{(i,j)}$ with $i\le j$ only,
because making replacements $\buc\leftrightarrow\bub$ and $\bvc\leftrightarrow\bvb$ one can  recast the remaining form factors
(see \eqref{F31-13}--\eqref{F21-12}). However,
the matrices $\mathcal{L}_{j,k}$ and $\mathcal{M}_{j,k}$, as well as the prefactor $H$ are not symmetric over these replacements.
Therefore, changing $\buc\leftrightarrow\bub$ and $\bvc\leftrightarrow\bvb$ in the determinant formulas given below we obtain
 more representations for the universal form factors.

\begin{itemize}
\item  We start with the diagonal form factors $\mathfrak{F}^{(i,i)}$. In this case $a'=a$ and $b'=b$.

Let $p$ be an integer from the set $\{1,\dots,a+b\}$, such that  $\Omega_p\ne 0$.
Then the universal form factors $\mathfrak{F}^{(i,i)}$ have the following determinant representations \cite{HutLPRS16c}:
\begin{equation}\label{Fii-det}
\mathfrak{F}^{(i,i)}\left(\begin{smallmatrix}
\buc & \bub \\ \bvc & \bvb \end{smallmatrix}\right)^{a,a}_{b,b}=\frac{H}{\Omega_p}  \; \det_{a+b} \mathcal{N}^{(i,i)}.
\end{equation}
The matrix elements $\mathcal{N}^{(i,i)}_{p,k}$  in the $p$-th row have the form
 \be{Yk}
 \begin{aligned}
&  \mathcal{N}^{(1,1)}_{p,k}=1+\frac{g(\bvb,x_k)}{g(\bvc,x_k)}-  \frac{f(\bvb,x_k)}{f(\bvc,x_k)}-\frac{f(x_k,\buc)}{f(x_k,\bub)}, \\
&  \mathcal{N}^{(2,2)}_{p,k}=-1,\\
&  \mathcal{N}^{(3,3)}_{p,k}= \mathcal{N}^{(1,1)}_{p,k}+\mathcal{N}^{(2,2)}_{p,k},
 \end{aligned}
 \qquad\qquad k=1,\dots,a+b.
 \ee
where $\bar x=\{\ub_1,\dots,\ub_a,\vc_1,\dots,\vc_{b}\}$. In the other rows the entries $\mathcal{N}^{(i,i)}_{j,k}$ do not
depend on $i$ and have the following form:
\be{MatMii}
\begin{aligned}
&\mathcal{N}^{(i,i)}_{j,k}=\mathcal{L}_{j,k},\qquad &j=1,\dots, a,\qquad &j\ne p,\\
&\mathcal{N}^{(i,i)}_{j+a,k}=\mathcal{M}_{j,k},\qquad &j=1,\dots, b, \qquad &j+a\ne p,
\end{aligned}
\ee
and $k=1,\dots,a+b$.
\end{itemize}

This determinant representation was obtained in \cite{HutLPRS16c}.
Note that the form factors are symmetric with respect to  any of the four sets of Bethe parameters. This symmetry follows from the
symmetry of the Bethe vectors.  Therefore, without any loss of generality one can  assume in \eqref{Fii-det}
that  $p=a$ or $p=a+b$.

\begin{itemize}

\item  For the universal form factor $\mathfrak{F}^{(1,2)}$, we have $a'=a+1$ and $b'=b$.
Let  $\Omega_{a+1}\ne 0$.   Then  $\mathfrak{F}^{(1,2)}$ has the form
\begin{equation}\label{UFF-12}
\mathfrak{F}^{(1,2)}\left(\begin{smallmatrix}
\buc & \bub \\ \bvc & \bvb \end{smallmatrix}\right)^{a+1,a}_{b,b}
=  \frac{H}{\Omega_{a+1}}  \; \det_{a+b} \mathcal{N}^{(1,2)},
\end{equation}
where
\be{MatM12}
\begin{aligned}
&\mathcal{N}^{(1,2)}_{j,k}=\mathcal{L}_{j,k},\qquad j=1,\dots, a,\\
&\mathcal{N}^{(1,2)}_{j+a,k}=\mathcal{M}_{j,k},\qquad j=1,\dots, b,
\end{aligned}
\ee
and $k=1,\dots,a+b$. The set $\bar x=\{\ub_1,\dots,\ub_a,\vc_1,\dots,\vc_{b}\}$.

\item  For the universal form factor $\mathfrak{F}^{(2,3)}$, we notice that $a'=a$ and $b'=b+1$.  Let  $\Omega_{a+b+1}\ne 0$.
Then  $\mathfrak{F}^{(2,3)}$ has the form
\begin{equation}\label{UFF-23}
\mathfrak{F}^{(2,3)}\left(\begin{smallmatrix}
\buc & \bub \\ \bvc & \bvb \end{smallmatrix}\right)^{a,a}_{b+1,b}
= (-1)^{b+1} \frac{H}{\Omega_{a+b+1}}  \;  \det_{a+b+1} \mathcal{N}^{(2,3)},
\end{equation}
where
\be{MatM23}
\begin{aligned}
&\mathcal{N}^{(2,3)}_{j,k}=\mathcal{L}_{j,k},\qquad j=1,\dots, a,\\
&\mathcal{N}^{(2,3)}_{j+a,k}=\mathcal{M}_{j,k},\qquad j=1,\dots, b,\\
&\mathcal{N}^{(2,3)}_{a+b+1,k}=1,
\end{aligned}
\ee
and $k=1,\dots,a+b+1$. The set $\bar x=\{\ub_1,\dots,\ub_a,\vc_1,\dots,\vc_{b+1}\}$.

\item  For the universal form factor $\mathfrak{F}^{(1,3)}$, one sees that $a'=a+1$ and $b'=b+1$.
Let  $\Omega_{a+1}\ne 0$. Then
 $\mathfrak{F}^{(1,3)}$ has the form
\begin{equation}\label{UFF-13a}
\mathfrak{F}^{(1,3)}\left(\begin{smallmatrix}
\buc & \bub \\ \bvc & \bvb \end{smallmatrix}\right)^{a+1,a}_{b+1,b} =
(-1)^{b+1}\frac{H}{\Omega_{a+1}} \;  \det_{a+b+1} \mathcal{N}^{(1,3)},
\end{equation}
where
\be{MatM13a}
\begin{aligned}
&\mathcal{N}^{(1,3)}_{j,k}=\mathcal{L}_{j,k},\qquad j=1,\dots, a,\\
&\mathcal{N}^{(1,3)}_{j+a,k}=\mathcal{M}_{j,k},\qquad j=1,\dots, b+1,
\end{aligned}
\ee
and $k=1,\dots,a+b+1$. The set $\bar x=\{\ub_1,\dots,\ub_a,\vc_1,\dots,\vc_{b+1}\}$.

\item   For the universal form factor $\mathfrak{F}^{(2,1)}$, one has $a'=a-1$ and $b'=b$. It has the form
\begin{equation}\label{A-UFF-12}
\mathfrak{F}^{(2,1)}\left(\begin{smallmatrix}
\buc & \bub \\ \bvc & \bvb \end{smallmatrix}\right)^{a-1,a}_{b,b}
=  H \det_{a+b} \mathcal{N}^{(2,1)},
\end{equation}
where
\be{A-MatM12}
\begin{aligned}
&\mathcal{N}^{(2,1)}_{j,k}=\mathcal{L}_{j,k},\qquad j=1,\dots, a-1,\\
&\mathcal{N}^{(2,1)}_{a,k}=-1,\\
&\mathcal{N}^{(2,1)}_{j+a,k}=\mathcal{M}_{j,k},\qquad j=1,\dots, b,
\end{aligned}
\ee
and $k=1,\dots,a+b$. The set $\bar x=\{\ub_1,\dots,\ub_a,\vc_1,\dots,\vc_{b}\}$.

\item  For the universal form factor $\mathfrak{F}^{(3,2)}$ with $a'=a$ and $b'=b-1$. It has the form
\begin{equation}\label{A-UFF-23}
\mathfrak{F}^{(3,2)}\left(\begin{smallmatrix}
\buc & \bub \\ \bvc & \bvb \end{smallmatrix}\right)^{a,a}_{b-1,b}
= (-1)^{b-1}H  \det_{a+b-1} \mathcal{N}^{(3,2)},
\end{equation}
where
\be{A-MatM23}
\begin{aligned}
&\mathcal{N}^{(3,2)}_{j,k}=\mathcal{L}_{j,k},\qquad j=1,\dots, a,\\
&\mathcal{N}^{(3,2)}_{j+a,k}=\mathcal{M}_{j,k},\qquad j=1,\dots, b-1,
\end{aligned}
\ee
and $k=1,\dots,a+b-1$. The set $\bar x=\{\ub_1,\dots,\ub_a,\vc_1,\dots,\vc_{b-1}\}$.

\item For the universal form factor $\mathfrak{F}^{(3,1)}$ and $a'=a-1$ and $b'=b-1$. It has the form
\begin{equation}\label{A-UFF-13a}
\mathfrak{F}^{(3,1)}\left(\begin{smallmatrix}
\buc & \bub \\ \bvc & \bvb \end{smallmatrix}\right)^{a-1,a}_{b-1,b} =
(-1)^{b-1}H  \det_{a+b-1} \mathcal{N}^{(3,1)},
\end{equation}
where
\be{A-MatM13a}
\begin{aligned}
&\mathcal{N}^{(3,1)}_{j,k}=\mathcal{L}_{j,k},\qquad j=1,\dots,  a-1,\\
&\mathcal{N}^{(3,1)}_{j+a,k}=\mathcal{M}_{j,k},\qquad j=1,\dots, b-1,\\
&\mathcal{N}^{(3,1)}_{a,k}=\frac{(-1)^{a-1}r_1(x_k)h(\buc,x_k)}{f(\bvc,x_k)h(x_k,\bub)}
 - \frac{h(x_k,\buc)}{h(x_k,\bub)},
\end{aligned}
\ee
and $k=1,\dots,a+b-1$. The set $\bar x=\{\ub_1,\dots,\ub_a,\vc_1,\dots,\vc_{b-1}\}$.

\end{itemize}

The proofs of the determinant representations for the universal form factors of off-diagonal matrix elements will be
given in section~\ref{S-detFF}.

\section{Proof of determinant formula \eqref{FF-res1}\label{S-PSS}}

Form factors of the operators $T_{ii}(z)$ with respect to identical states were calculated in \cite{HutLPRS16c}.
There it was shown that $\mathcal{F}^{(i,i)}$ are proportional to the $\kappa_i$-derivative of the twisted transfer matrix eigenvalue:
\be{FF-res-new1}
\mathcal{F}^{(i,i)}\left(z\Bigl|\begin{smallmatrix}
\bu & \bu \\ \bv & \bv \end{smallmatrix}\right)^{a,a}_{b,b}=(-1)^{[i]}\frac{d\tau_\kappa(z|\buc,\bvc)}{d\kappa_i}\Bigr|_{\bar\kappa=1}
\mathbb{C}_{a,b}(\bu;\bv)\mathbb{B}_{a,b}(\bu;\bv).
\ee
A peculiarity of this representation is that we have a {\it full} derivative  of $\tau_\kappa(z|\buc,\bvc)$ over $\kappa_i$. In other words,
one should consider the Bethe parameters $\buc$ and $\bvc$ as implicit functions of $\kappa_i$, whose dependence on the twist parameters is
determined by the twisted Bethe equations \eqref{K-AEigenS-1}. In this section we show that representation \eqref{FF-res-new1} and \eqref{FF-res1}
are equivalent.

Consider a solution $\{\buc(\kappa),\bvc(\kappa)\}$ of the twisted Bethe equations such that $\{\buc(\kappa),\bvc(\kappa)\}\to \{\bu,\bv\}$ as
$\bar \kappa\to 1$. Then, similarly to \eqref{Phi}, we introduce an $(a+b)$-component vector $\Phic$ as
\be{Twist-Phi}
\begin{aligned}
\Phic_j&=\log\left(\frac{r_1(\uc_j)}{f(\bvc,\uc_j)}\frac{f(\buc_j,\uc_j)}{f(\uc_j,\buc_j)}\right),\qquad j=1,\dots,a,\\
\Phic_{a+j}&=\log\left(\frac{r_3(\vc_j)}{f(\vc_j,\buc)}\right),\qquad j=1,\dots,b.
\end{aligned}
\ee
Comparing this vector with the vector $\Phi$ \eqref{Phi} we see that $\Phic\to\Phi$ as\footnote{Here and below $\bar \kappa=1$  stands for $\kappa_1=\kappa_2=\kappa_3=1$.  We also assume that the condition $\bar\kappa=1$ automatically yields $\buc=\bu$ and $\bvc=\bv$.} $\bar\kappa\to 1$.

Taking the logarithm of the twisted Bethe equations \eqref{K-AEigenS-1} we obtain
\be{Log-TBE}
\begin{aligned}
&\Phic_j=\log\left(\frac{\kappa_2}{\kappa_1}\right),\qquad j=1,\dots,a,\\
&\Phic_{a+j}=\log\left(\frac{\kappa_2}{\kappa_3}\right),\qquad j=1,\dots,b.
\end{aligned}
\ee
Differentiating these equations over $\kappa_i$ at
$\bar \kappa=1$ 
we find
\be{Log-TBEa}
\begin{aligned}
&\sum_{k=1}^a\frac{\partial\Phi_j}{\partial u_k}\frac{d\uc_k}{d\kappa_i}\Bigr|_{\bar\kappa=1}+
\sum_{k=1}^b\frac{\partial\Phi_j}{\partial v_k}\frac{d\vc_k}{d\kappa_i}\Bigr|_{\bar\kappa=1}
=\delta_{2i}-\delta_{1i},\qquad j=1,\dots,a,\\
&\sum_{k=1}^a\frac{\partial\Phi_{a+j}}{\partial u_k}\frac{d\uc_k}{d\kappa_i}\Bigr|_{\bar\kappa=1}+
\sum_{k=1}^b\frac{\partial\Phi_{a+j}}{\partial v_k}\frac{d\vc_k}{d\kappa_i}\Bigr|_{\bar\kappa=1}
=\delta_{2i}-\delta_{3i},\qquad j=1,\dots,b,
\end{aligned}
\ee
where we have taken into account that $\Phic_j=\Phi_j$, $\uc_j=u_j$, and $\vc_j=v_j$ at $\bar\kappa=1$.

Let $\bar x=\{\uc_1,\dots,\uc_a,\vc_1,\dots,\vc_b\}$. Then using \eqref{hatNii} we recast \eqref{Log-TBEa} as follows:
\be{Log-TBE-N}
\begin{aligned}
&\sum_{k=1}^{a+b} \widehat{\mathcal{N}}^{(i,i)}_{j,k}\frac{dx_k}{d\kappa_i}\Bigr|_{\bar\kappa=1}
=\delta_{2i}-\delta_{1i},\qquad j=1,\dots,a,\\
&\sum_{k=1}^{a+b} \widehat{\mathcal{N}}^{(i,i)}_{a+j,k}\frac{dx_k}{d\kappa_i}\Bigr|_{\bar\kappa=1}
=\delta_{2i}-\delta_{3i},\qquad j=1,\dots,b.
\end{aligned}
\ee
Hence, if we multiply the columns  $\widehat{\mathcal{N}}^{(i,i)}_{j,k}$ with $k=1,\dots,a+b$ by the coefficients $dx_k/d\kappa_i$ and add this linear combination to the last column of the matrix $\widehat{\mathcal{N}}^{(i,i)}$, then we obtain zeros everywhere except the right-lower element. For this non-zero entry we obtain
\begin{multline}\label{Nonzero}
\widehat{\mathcal{N}}^{(i,i)}_{a+b+1,a+b+1}+\sum_{k=1}^{a+b} \widehat{\mathcal{N}}^{(i,i)}_{a+b+1,k}\frac{dx_k}{d\kappa_i}\Bigr|_{\bar\kappa=1}\\
=(-1)^{[i]}\frac{\partial\tau_\kappa(z|\bu,\bv)}{\partial \kappa_i}+ (-1)^{[i]}\sum_{k=1}^{a+b}  \frac{\partial\tau(z|\buc,\bvc)}{\partial x_k}
\frac{dx_k}{d\kappa_i}\Bigr|_{\bar\kappa=1}=(-1)^{[i]}\frac{d\tau_\kappa(z|\buc,\bvc)}{d\kappa_i}\Bigr|_{\bar\kappa=1}.
\end{multline}
Thus, we arrive at
\begin{multline}\label{FF-res-new}
\mathcal{F}^{(i,i)}\left(z\Bigl|\begin{smallmatrix}
\bu & \bu \\ \bv & \bv \end{smallmatrix}\right)^{a,a}_{b,b}=(-c)^{a+b}\prod_{j=1}^b\prod_{k=1}^a f(v_j,u_k)\prod_{\substack{j,k=1\\j\ne k}}^a f(u_j,u_k)\;
\prod_{\substack{j,k=1\\j\ne k}}^b g(v_j,v_k)\\
\times (-1)^{[i]}\frac{d\tau_\kappa(z|\buc,\bvc)}{d\kappa_i}\Bigr|_{\bar\kappa=1}\det_{a+b}\widehat{\mathcal{N}}^{(i,i)},
\end{multline}
where now the size of the matrix $\widehat{\mathcal{N}}^{(i,i)}$ is $(a+b)\times(a+b)$. Comparing this expression with \eqref{Norm-res} we reproduce
representation \eqref{FF-res-new1}. \qed

\section{Zero modes\label{S-ZM}}

We  have shown in the paper \cite{PakRS15a} that in the models with $\mathfrak{gl}(N)$-invariant $R$-matrix all the form factors can be
obtained  from  one initial form  factor and taking special limits of the Bethe parameters. Our method was based on the
use of zero modes of the monodromy matrix. This approach  can be applied to the models with $\mathfrak{gl}(m|n)$
symmetry without significant changes. In this section we give a brief description of this method and find simple relations between different form factors.

The basis of the zero modes method is an expansion of the  monodromy matrix $T(u)$   into a series over inverse spectral parameter $u^{-1}$
\be{zero-modes}
T_{ij}(u)=\delta_{ij}\mathbf{1}+ \sum_{n=0}^\infty T_{ij}[n]\,\left(\tfrac cu\right)^{n+1}.
\ee
This expansion is typical if the monodromy matrix of the model is obtained as specialization to some highest weight representation
of the Yangian   $Y(\mathfrak{gl}(2|1))$ with highest weight vector $\vac$ \cite{Dr88,Molevbook}.

Note that the expansion \eqref{zero-modes} yields similar expansions for the functions $\lambda_i(u)$ and $r_k(u)$
\be{zero-modes-r}
\begin{aligned}
\lambda_i(u)&=1+ \sum_{n=0}^\infty \lambda_i[n]\,\left(\tfrac cu\right)^{n+1},\qquad i=1,2,3\\
r_k(u)&=1+ \sum_{n=0}^\infty r_k[n]\,\left(\tfrac cu\right)^{n+1},\qquad k=1,3.
\end{aligned}
\ee
Assumption \eqref{zero-modes} implies that the Bethe vectors remain on-shell if one of their parameters
tends to infinity. This is because the structure of the Bethe equations \eqref{AEigenS-1} is preserved
 when $r_k(u)\to 1$ at $u\to\infty$.

The operators $T_{ij}[0]$ are called the zero modes.  They span a $\mathfrak{gl}(2|1)$ superalgebra.
Sending in \eqref{RTT-comp} one of the arguments to infinity we obtain commutation relations of the zero modes and
the operators $T_{kl}(z)$
\be{RTT-zero}
 [T_{ij}[0],T_{kl}(z)\} = (-1)^{[l]([i]+[j])+[i][j]}\bigl(\delta_{il} T_{kj}(z) - \delta_{kj} T_{il}(z)\bigr),
\ee
 showing that the monodromy entries form an adjoint representation of the $\mathfrak{gl}(2|1)$ superalgebra
generated by the zero modes.

\subsection{Action of the  zero modes onto Bethe vectors}

The explicit formulas for the action the operators $T_{ij}(z)$ onto Bethe vectors were derived in \cite{HutLPRS16a}. Taking
the limit $z\to\infty$ in those expressions we obtain the action of the zero modes $T_{ij}[0]$. The action of $T_{ij}[0]$ with
$i<j$ is given by
\begin{align}
T_{13}[0]\mathbb{B}_{a,b}(\bu;\bv)&=
-\lim_{w\to\infty} \left(-\tfrac wc\right)^{b+1}\; \mathbb{B}_{a+1,b+1}(\{\bu,w\};\{\bv,w\}),\label{act-13}\\
T_{23}[0]\mathbb{B}_{a,b}(\bu;\bv)&=
-\lim_{w\to\infty} \left(-\tfrac wc\right)^{b+1}\; \mathbb{B}_{a,b+1}(\bu;\{\bv,w\}),\label{act-23}\\
T_{12}[0]\mathbb{B}_{a,b}(\bu;\bv)&=
\lim_{w\to\infty} \tfrac wc\; \mathbb{B}_{a+1,b}(\{\bu,w\};\bv).
\label{act-12}
\end{align}

Let us show how one can obtain these equations. For this we consider the simplest case \eqref{act-13}. The action of the
operator $T_{13}(w)$ onto a Bethe vector $\mathbb{B}_{a,b}(\bu;\bv)$ is (see \cite{HutLPRS16a})
\be{Deract-131}
T_{13}(w)\mathbb{B}_{a,b}(\bu;\bv)=\lambda_2(w)
h(\bv, w)\; \mathbb{B}_{a+1,b+1}(\{\bu,w\};\{\bv,w\}).
\ee
Multiplying both sides by $w/c$, taking the limit $w\to\infty$, and using the asymptotic properties of the  functions $h(v,w)$
\eqref{propert} and $\lambda_2(w)$ \eqref{zero-modes-r}   we immediately arrive at \eqref{act-13}.

The parameters $\bu$ and $\bv$ in \eqref{act-13}--\eqref{act-12} are a priori generic
complex numbers, but they may satisfy  the Bethe equations in specific cases. Then in the r.h.s. of \eqref{act-23} and \eqref{act-12} we
obtain on-shell Bethe vectors, because the infinite root $w$ together with the sets $\bu$ and $\bv$ satisfy Bethe equations due to the
condition \eqref{zero-modes-r}.

Applying the antimorphism $\psi$ to the actions \eqref{act-13}--\eqref{act-12} we obtain
\begin{align}
\mathbb{C}_{a,b}(\bu;\bv)T_{31}[0]&=\lim_{w\to\infty} \left(\tfrac wc \right)^{b+1}\mathbb{C}_{a+1,b+1}(\{\bu,w\};\{\bv,w\}),\label{ZM-dBV31}\\
\mathbb{C}_{a,b}(\bu;\bv)T_{32}[0]&=\lim_{w\to\infty} \left(\tfrac wc \right)^{b+1}\mathbb{C}_{a+1,b}(\bu;\{\bv,w\}),\label{ZM-dBV32}\\
\mathbb{C}_{a,b}(\bu;\bv)T_{21}[0]&=\lim_{w\to\infty} \tfrac wc \;\mathbb{C}_{a+1,b}(\{\bu,w\};\bv).\label{ZM-dBV21}
\end{align}
As in the above case, if the parameters $\{\bu,\bv\}$ satisfy Bethe equations, then $\{\bu,\bv,w\}$ also satisfy Bethe equations as
$w\to\infty$.

Similarly to the $\mathfrak{gl}(N)$ case (see  \cite{MuhTV06})
the on-shell vectors (resp. dual on-shell vectors) depending on finite Bethe roots are  {\it singular weight} vectors of the zero modes  $T_{ij}[0]$
with $i>j$ (resp. $T_{ij}[0]$ with $i<j$):
\be{sing-vect}
\begin{aligned}
T_{ij}[0]\mathbb{B}_{a,b}(\bu;\bv)&=0,\qquad i>j,\\
\mathbb{C}_{a,b}(\bu;\bv)T_{ij}[0]&=0,\qquad i<j.
\end{aligned}
\ee
These equations can be obtained from the explicit formulas of the actions of $T_{ij}$ onto Bethe vectors \cite{HutLPRS16a}.

\subsection{Relations between different form factors}

The zero modes allow us to find simple relations between different form factors.  As a starter, we consider an
example. Setting in \eqref{RTT-zero} $j=k=l=2$ and $i=1$ we obtain
\be{RTT-exp}
[T_{12}[0],T_{22}(z)]=-T_{12}(z).
\ee

Let $\mathbb{C}_{a+1,b}(\buc;\bvc)$ and $\mathbb{B}_{a,b}(\bub;\bvb)$ be two on-shell vectors with
all  their Bethe parameters finite.  Then \eqref{RTT-exp} yields
\begin{multline}
\mathbb{C}_{a+1,b}(\buc;\bvc) T_{12}(z)\mathbb{B}_{a,b}(\bub;\bvb)=
-\mathbb{C}_{a+1,b}(\buc;\bvc) T_{12}[0]T_{22}(z)\mathbb{B}_{a,b}(\bub;\bvb) \\
+\mathbb{C}_{a+1,b}(\buc;\bvc) T_{22}(z)T_{12}[0]\mathbb{B}_{a,b}(\bub;\bvb) .\label{T13-z1}
\end{multline}
The first term in the r.h.s. vanishes as $T_{12}[0]$ acts on the dual on-shell Bethe vector. The action of
$T_{12}[0]$ on the on-shell vector $\mathbb{B}_{a,b}(\bub;\bvb)$  is given by \eqref{act-12}, hence,
\begin{equation}
\mathbb{C}_{a+1,b}(\buc;\bvc) T_{12}(z)\mathbb{B}_{a,b}(\bub;\bvb)=
\mathbb{C}_{a+1,b}(\buc;\bvc) T_{22}(z)\lim_{w\to\infty}\tfrac{w}c\;\mathbb{B}_{a+1,b}(\{\bub,w\};\bvb) .\label{T13-z2}
\end{equation}
Since the original vector $\mathbb{B}_{a,b}(\bub;\bvb)$ was on-shell, the new vector  $\mathbb{B}_{a+1,b}(\{\bub,w\};\bvb)$ with $w\to\infty$ also is on-shell. Therefore, in the r.h.s.
of \eqref{T13-z2} we have the form factor of $T_{22}(z)$, and we arrive at
\begin{equation}\label{F12-ee0}
  \mathcal{F}^{(1,2)}\left(z\Bigl|\begin{smallmatrix}\buc & \bub \\ \bvc & \bvb
\end{smallmatrix}\right)^{a+1,a}_{b,b}= \lim_{w\to\infty} \frac{w}{c}\mathcal{F}^{(2,2)}\left(z\Bigl|\begin{smallmatrix}
\buc & \{\bub,w\} \\ \bvc & \bvb \end{smallmatrix}\right)^{a+1,a+1}_{b,b}.
\end{equation}
Thus, the form factor $\mathcal{F}^{(1,2)}$ can be obtained from $\mathcal{F}^{(2,2)}$ by sending one of the Bethe parameters to infinity.

The relation \eqref{F12-ee0} can be easily reformulated for the universal form factors. Indeed, looking at the explicit expression \eqref{tau-def} for the eigenvalue $\tau(z|\bu,\bv)$ we see that
\be{lim-tau}
\lim_{u_j\to\infty}\tau(z|\bu,\bv)=\tau(z|\bu_j,\bv),\qquad
\lim_{v_k\to\infty}\tau(z|\bu,\bv)=\tau(z|\bu,\bv_k).
\ee
Thus, if one of the Bethe parameters goes to infinity, then the transfer matrix eigenvalue $\tau(z|\bu,\bv)$ turns into the
eigenvalue depending on the remaining Bethe parameters. Hence, we arrive at
\begin{equation}\label{F12-ee}
  \mathfrak{F}^{(1,2)}\left(\begin{smallmatrix}\buc & \bub \\ \bvc & \bvb
\end{smallmatrix}\right)^{a+1,a}_{b,b}= \lim_{w\to\infty} \frac{w}{c}\,\mathfrak{F}^{(2,2)}\left(\begin{smallmatrix}
\buc & \{\bub,w\} \\ \bvc & \bvb \end{smallmatrix}\right)^{a+1,a+1}_{b,b}.
\end{equation}

Similarly, starting with the universal form factor $\mathfrak{F}^{(2,2)}$ and using commutation relations \eqref{RTT-zero}
we can obtain all the universal form factors  $\mathfrak{F}^{(i,j)}$ with $|i-j|=1$:
\begin{equation}\label{F23-ee}
\mathfrak{F}^{(2,3)}\left(\begin{smallmatrix}
\buc & \bub \\ \bvc & \bvb \end{smallmatrix}\right)^{a,a}_{b+1,b}=\lim_{w\to\infty} \left(\frac{-w}{c}\right)^{b+1}
\mathfrak{F}^{(2,2)}\left(\begin{smallmatrix}
\buc & \bub \\ \bvc & \{\bvb,w\} \end{smallmatrix}\right)^{a,a}_{b+1,b+1},
\end{equation}
\begin{equation}\label{F21-ee}
  \mathfrak{F}^{(2,1)}\left(\begin{smallmatrix}\buc & \bub \\ \bvc & \bvb
\end{smallmatrix}\right)^{a-1,a}_{b,b}= \lim_{w\to\infty} \frac{w}{c}\,\mathfrak{F}^{(2,2)}\left(\begin{smallmatrix}
\{\buc,w\} & \bub \\ \bvc & \bvb \end{smallmatrix}\right)^{a,a}_{b,b},
\end{equation}
\begin{equation}\label{F32-ee}
\mathfrak{F}^{(3,2)}\left(\begin{smallmatrix}
\buc & \bub \\ \bvc & \bvb \end{smallmatrix}\right)^{a,a}_{b-1,b}=-\lim_{w\to\infty} \left(\frac{w}{c}\right)^{b}
\mathfrak{F}^{(2,2)}\left(\begin{smallmatrix}
\buc & \bub \\ \{\bvc,w\} & \bvb \end{smallmatrix}\right)^{a,a}_{b,b}.
\end{equation}
The universal form factors  $\mathfrak{F}^{(i,j)}$ with $|i-j|=2$  can be obtained as the limits of $\mathfrak{F}^{(i,j)}$
with $|i-j|=1$, for example,
\begin{equation}\label{F13-ee}
  \mathfrak{F}^{(1,3)}\left(\begin{smallmatrix}\buc & \bub \\ \bvc & \bvb
\end{smallmatrix}\right)^{a+1,a}_{b+1,b}= \lim_{w\to\infty} \left(\frac{-w}{c}\right)^{b+1}\mathfrak{F}^{(1,2)}\left(\begin{smallmatrix}
\buc & \bub \\ \bvc & \{\bvb,w\} \end{smallmatrix}\right)^{a+1,a}_{b+1,b+1},
\end{equation}
\begin{equation}\label{F31-ee}
\mathfrak{F}^{(3,1)}\left(\begin{smallmatrix}
\buc & \bub \\ \bvc & \bvb \end{smallmatrix}\right)^{a-1,a}_{b-1,b}=\lim_{w\to\infty} \frac{w}{c}
\mathfrak{F}^{(3,2)}\left(\begin{smallmatrix}
\{\buc,w\} & \bub \\ \bvc & \bvb \end{smallmatrix}\right)^{a,a}_{b-1,b}.
\end{equation}
Thus, starting with $\mathfrak{F}^{(2,2)}$ and taking different limits of the Bethe parameters we obtain all the universal form factors of
the off-diagonal matrix elements of the monodromy matrix. Formally, $\mathfrak{F}^{(1,1)}$ and $\mathfrak{F}^{(3,3)}$ can be also included in this
scheme, for example,
\be{F11F22-F12}
\mathfrak{F}^{(1,1)}\left(\begin{smallmatrix}
\buc & \bub \\ \bvc & \bvb \end{smallmatrix}\right)^{a,a}_{b,b}-
\mathfrak{F}^{(2,2)}\left(\begin{smallmatrix}
\buc & \bub \\ \bvc & \bvb \end{smallmatrix}\right)^{a,a}_{b,b}
=\lim_{w\to\infty}\frac{w}c\; \mathfrak{F}^{(1,2)}\left(\begin{smallmatrix}
\{\buc,w\} & \bub \\ \bvc & \bvb \end{smallmatrix}\right)^{a+1,a}_{b,b}.
\ee
However, in our case this relation is not needed, because we have already determinant representations for all diagonal universal
form factors \cite{HutLPRS16c}.

It should be noted that the possibility of considering the limit of an infinite Bethe parameter is based on the use of the generalized model. On the one hand, in this model, the Bethe parameters  are arbitrary complex numbers. Hence, one of them can be sent to infinity. On the other
hand, the existence  of an infinite root in the Bethe equations agrees  with the expansion \eqref{zero-modes-r}.
At the same time, since the final expression for form factors depends on $r_1$ and $r_3$ only through the eigenvalue
$\tau(z|\bu,\bv)$, the condition \eqref{zero-modes-r} is not a restriction on the form factors. It can be checked for instance in Bose gas models \cite{PakRS15b}, where the relations between for factors and the zero modes method both apply, although the condition \eqref{zero-modes-r} is not fulfilled.

\section{Form factors of off-diagonal elements\label{S-detFF}.}

In this section we  deduce from the zero modes method  determinant representations for the universal  form factors of the operators $T_{ij}(z)$
with $i\ne j$. We restrict ourselves with two typical examples of $\mathfrak{F}^{(1,2)}$ and $\mathfrak{F}^{(3,2)}$. All other determinant
representations for the universal form factors can be obtained in a similar manner.

\subsection{Form factor $\mathfrak{F}^{(1,2)}$}

Due to \eqref{F12-ee} the form factor $\mathfrak{F}^{(1,2)}$ is a limiting case of the form factor $\mathfrak{F}^{(2,2)}$. The determinant
representation for the latter is given by \eqref{Fii-det}--\eqref{MatMii}, where without any loss of generality we can set $p=a+1$.
In these expressions we also should replace $a$ with $a+1$ and $\bub$ with $\{\bub,w\}$
Then we have
\begin{equation}\label{F11-22a}
\mathfrak{F}^{(1,2)}\left(\begin{smallmatrix}
\buc & \bub \\ \bvc & \bvb \end{smallmatrix}\right)^{a+1,a}_{b,b}=\lim_{w\to\infty}\frac{w H}{c\, \Omega_{a+1}}   \det_{a+b+1} \mathcal{N}^{(2,2)}.
\end{equation}
For taking the limit it is convenient to multiply the first $a$ rows of the matrix $\mathcal{N}^{(2,2)}$ by the factors  $-w/c$. Then we obtain
\begin{equation}\label{F11-22b}
\mathfrak{F}^{(1,2)}\left(\begin{smallmatrix}
\buc & \bub \\ \bvc & \bvb \end{smallmatrix}\right)^{a+1,a}_{b,b}=\lim_{w\to\infty}
\left(\frac{-c}w\right)^a  \frac{w H}{c\, \Omega_{a+1}}   \det_{a+b+1}\circN_{j,k}^{(2,2)}.
\end{equation}
where
\be{circN22}
\begin{aligned}
&\circN_{j,k}^{(2,2)}=- \frac wc  {\mathcal{N}}{}_{j,k}^{(2,2)},\qquad j=1,\dots,a,\\
&\circN_{j,k}^{(2,2)}={\mathcal{N}}{}_{j,k}^{(2,2)},\qquad j=a+1,\dots,a+b+1.
\end{aligned}
\ee

Now let us give explicit expressions for the prefactor and the matrix elements in \eqref{F11-22b}. The factor
$H$ is
\begin{multline}\label{Hab1}
H(\buc;\{\bub,w\};\bvc)= f(\bvc,\bub)h(\bub,\bub)\Delta'(\buc)\Delta(\bub)\Delta(\bvc)\Delta'(\bvc)\\
\times f(\bvc,w)h(w,\bub)h(\bub,w)g(w,\bub)= f(\bvc,w)f(w,\bub)h(\bub,w)H(\buc;\bub;\bvc),
\end{multline}
where $H(\buc;\bub;\bvc)$ is given by \eqref{Hab0}. Hence, due to \eqref{propert} we find
\be{limH}
\lim_{w\to\infty}\left( \frac{-c}w  \right)^aH(\buc;\{\bub,w\};\bvc)=H(\buc;\bub;\bvc).
\ee
 The coefficient $\Omega_{a+1}$ is equal to
\be{vector-w}
\Omega_{a+1}(\buc;\{\bub,w\})=\frac1{g(\uc_{a+1},w)}\;\frac{g(\uc_{a+1},\buc_{a+1})}{g(\uc_{a+1},\bub)}=\frac{\Omega_{a+1}(\buc;\bub)}{g(\uc_{a+1},w)},
\ee
and therefore
\be{limXi}
\lim_{w\to\infty}\frac cw\;\Omega_{a+1}(\buc;\{\bub,w\})=-\Omega_{a+1}(\buc;\bub),
\ee
where $\Omega_{a+1}(\buc;\bub)$ is given by \eqref{vector}. Thus, the prefactor coincides with the one in \eqref{UFF-12}  up to the sign.

Consider now the matrix elements $\circN_{j,k}^{(2,2)}$. First of all $\circN_{a+1,k}^{(2,2)}=-1$ for all $k=1,\dots,a+b+1$.
If $j,k\ne a+1$, then
\begin{multline}\label{matMup-w}
 \circN_{j,k}^{(2,2)}(\buc;\{\bub,w\};\bvc) \\
 =-\frac wc \left(\frac{(-1)^{a}r_1(x_k)t(\uc_j,x_k)h(\buc,x_k)}{f(\bvc,x_k)h(x_k,\bub)h(x_k,w)}
 +  \frac{t(x_k,\uc_j)h(x_k,\buc)}{h(x_k,\bub)h(x_k,w)}\right),
 \qquad \begin{array}{l}j=1,\dots,a, \\
  k=1,\dots,  a+b+1,\\  k\ne a+1,\end{array}
\end{multline}
\begin{multline}\label{matMdown-w}
 \circN_{a+1+j,k}^{(2,2)}(\{\bub,w\};\bvc;\bvb)\\
  = - t(\vc_j,x_k) \frac{g(\bvb,x_k)}{g(\bvc,x_k)}\left(1-\frac{r_3(x_k)}{f(x_k,\bub)f(x_k,w)}\right), \qquad
\begin{array}{l}j=1,\dots,b, \\
k=1,\dots,  a+b+1,\\  k\ne a+1.\end{array}
\end{multline}
Here $\{x_1,\dots,x_a\}=\{\ub_1,\dots,\ub_a\}$ and $\{x_{a+2},\dots,x_{a+b+1}\}=\{\vc_1,\dots,\vc_b\}$. Taking the limit $w\to\infty$ we obtain
\begin{multline}\label{limmatMup-w}
\lim_{w\to\infty}\circN_{j,k}^{(2,2)}(\buc;\{\bub,w\};\bvc) \\
 = \frac{(-1)^{a}r_1(x_k)t(\uc_j,x_k)h(\buc,x_k)}{f(\bvc,x_k)h(x_k,\bub)}
 +  \frac{t(x_k,\uc_j)h(x_k,\buc)}{h(x_k,\bub)},
 \qquad \begin{array}{l}j=1,\dots,a, \\
  k=1,\dots,  a+b+1,\\  k\ne a+1,\end{array}
\end{multline}
\begin{multline}\label{limmatMdown-w}
\lim_{w\to\infty} \circN_{a+1+j,k}^{(2,2)}(\{\bub,w\};\bvc;\bvb)\\
  = - t(\vc_j,x_k) \frac{g(\bvb,x_k)}{g(\bvc,x_k)}\left(1-\frac{r_3(x_k)}{f(x_k,\bub)}\right), \qquad
\begin{array}{l}j=1,\dots,b, \\
k=1,\dots,  a+b+1,\\  k\ne a+1.\end{array}
\end{multline}
Finally, for the elements $\circN_{j,a+1}^{(2,2)}$ with $j\ne a+1$ we have
\begin{equation}\label{matMup-ww}
\circN_{j,a+1}^{(2,2)}(\buc;\{\bub,w\};\bvc) = \frac{-w}c \left( t(\uc_j,w)\frac{(-1)^{a}r_1(w)h(\buc,w)}{f(\bvc,w)h(w,\bub)}
 +  \frac{t(w,\uc_j)h(w,\buc)}{h(w,\bub)}\right),\quad  j<a+1,
\end{equation}
\begin{equation}\label{matMdown-ww}
\circN_{j,a+1}^{(2,2)}(\{\bub,w\};\bvc;\bvb) = - t(\vc_j,w) \frac{g(\bvb,w)}{g(\bvc,w)},\quad  j>a+1,
\end{equation}
and sending there $w$ to infinity we obtain
\be{limLM}
\lim_{w\to\infty} \circN_{j,a+1}^{(2,2)}(\buc;\{\bub,w\};\bvc)=
\lim_{w\to\infty} \circN_{j,a+1}^{(2,2)}(\{\bub,w\};\bvc;\bvb)=0.
\ee
We see that the $(a+1)$-th column of the matrix $\circN_{j,k}^{(2,2)}$ contains only one non-zero element
$\circN_{a+1,a+1}^{(2,2)}=-1$. Thus, the determinant in \eqref{F11-22b} reduces to the determinant of the $(a+b)\times(a+b)$
matrix with the matrix elements \eqref{limmatMup-w} and \eqref{limmatMdown-w}. Obviously, this representation coincides with
the expressions \eqref{UFF-12} and \eqref{MatM12}.

\subsection{Form factor $\mathfrak{F}^{(3,2)}$}

The form factor $\mathfrak{F}^{(3,2)}$ also can be obtained as a limit of the form factor $\mathfrak{F}^{(2,2)}$ via \eqref{F32-ee}. We
use again representation  \eqref{Fii-det}--\eqref{MatMii}, but now it is convenient to set $p=a+b$. We also should replace $\bvc$ with $\{\bvc,w\}$.
Then
\begin{equation}\label{A-F32-ee}
\mathfrak{F}^{(3,2)}\left(\begin{smallmatrix}
\buc & \bub \\ \bvc & \bvb \end{smallmatrix}\right)^{a,a}_{b-1,b}=-\lim_{w\to\infty} \left(\frac{w}{c}\right)^{b}
 \frac{H}{\Omega_{a+b}}  \; \det_{a+b} \mathcal{N}^{(2,2)}.
\end{equation}
For taking the limit we multiply the  rows with $j=a+1,\dots,a+b-1$ of the matrix $\mathcal{N}^{(2,2)}$ by the factors $c/w$. Then we obtain
\begin{equation}\label{F32-22b}
\mathfrak{F}^{(3,2)} \left(\begin{smallmatrix}
\buc & \bub \\ \bvc & \bvb \end{smallmatrix}\right)^{a,a}_{ b-1,b}  =-\lim_{w\to\infty}\frac wc
\left(\frac{w}c\right)^{2b-2}  \frac{H}{\Omega_{a+b}} \; \det_{a+b}\circN_{j,k}^{(2,2)}.
\end{equation}
where
\be{A-circN22}
\begin{aligned}
&\circN_{j,k}^{(2,2)}={\mathcal{N}}{}_{j,k}^{(2,2)},\qquad j=1,\dots,a,\\
&\circN_{j,k}^{(2,2)}=\frac cw{\mathcal{N}}{}_{j,k}^{(2,2)},\qquad j=a+1,\dots,a+b-1,\\
&\circN_{a+b,k}^{(2,2)}={\mathcal{N}}{}_{a+b,k}^{(2,2)}=-1.
\end{aligned}
\ee

Now let us give explicit expressions for the prefactor and the matrix elements in \eqref{F32-22b}. The factor
$H$ is
\begin{multline}\label{A-Hab1}
H(\buc;\bub;\{\bvc,w\})= f(\bvc,\bub)h(\bub,\bub)\Delta'(\buc)\Delta(\bub)\Delta(\bvc)\Delta'(\bvc)\\
\times f(w,\bub)g(w,\bvc)g(\bvc,w)= f(w,\bub)g(w,\bvc)g(\bvc,w)H(\buc;\bub;\bvc),
\end{multline}
where $H(\buc;\bub;\bvc)$ is given by \eqref{Hab0}. Hence, due to \eqref{propert} we find
\be{A-limH}
\lim_{w\to\infty}\left(\frac{w}c\right)^{2b-2}H(\buc;\bub;\{\bvc,w\})=(-1)^{b-1}H(\buc;\bub;\bvc).
\ee
 The coefficient $\Omega_{a+b}$ is equal to
\be{A-vector-w}
\Omega_{a+b}(\{\bvc,w\};\bub)=\frac{g(w,\bvc)}{g(w,\bvb)},
\ee
and therefore
\be{A-limXi}
\lim_{w\to\infty}\frac cw\;\Omega_{a+b}(\{\bvc,w\};\bub)=1.
\ee

Consider now the matrix elements $\circN_{j,k}^{(2,2)}$.
If  $k\ne a+b$, then
\begin{multline}\label{A-matMup-w}
 \circN_{j,k}^{(2,2)}(\buc;\bub;\{\bvc,w\}) \\
 =\frac{(-1)^{a-1}r_1(x_k)t(\uc_j,x_k)h(\buc,x_k)}{f(\bvc,x_k)f(w,x_k)h(x_k,\bub)}
 +  \frac{t(x_k,\uc_j)h(x_k,\buc)}{h(x_k,\bub)},
 \qquad \begin{array}{l}j=1,\dots,a, \\
  k=1,\dots, a+b-1,\end{array}
\end{multline}
\begin{multline}\label{A-matMdown-w}
 \circN_{a+j,k}^{(2,2)}(\bub;\{\bvc,w\};\bvb)\\
  = - \frac cw t(\vc_j,x_k) \frac{g(\bvb,x_k)}{g(\bvc,x_k)g(w,x_k)}\left(1-\frac{r_3(x_k)}{f(x_k,\bub)}\right), \qquad
\begin{array}{l}j=1,\dots,b-1, \\
k=1,\dots, a+b-1.\end{array}
\end{multline}
Here $\{x_1,\dots,x_a\}=\{\ub_1,\dots,\ub_a\}$ and $\{x_{a+1},\dots,x_{a+b-1}\}=\{\vc_1,\dots,\vc_{b-1}\}$. Taking the limit $w\to\infty$ we obtain
\begin{multline}\label{A-limmatMup-w}
\lim_{w\to\infty}\circN_{j,k}^{(2,2)}(\buc;\bub;\{\bvc,w\}) \\
 = \frac{(-1)^{a-1}r_1(x_k)t(\uc_j,x_k)h(\buc,x_k)}{f(\bvc,x_k)h(x_k,\bub)}
 +  \frac{t(x_k,\uc_j)h(x_k,\buc)}{h(x_k,\bub)},
 \qquad \begin{array}{l}j=1,\dots,a, \\
  k=1,\dots, a+b-1,\end{array}
\end{multline}
\begin{multline}\label{A-limmatMdown-w}
\lim_{w\to\infty} \circN_{a+j,k}^{(2,2)}(\bub;\{\bvc,w\};\bvb)\\
  = -  t(\vc_j,x_k) \frac{g(\bvb,x_k)}{g(\bvc,x_k)}\left(1-\frac{r_3(x_k)}{f(x_k,\bub)}\right), \qquad
\begin{array}{l}j=1,\dots,b, \\
k=1,\dots, a+b-1.\end{array}
\end{multline}
Finally, for the elements $\circN_{j,a+b}^{(2,2)}$ with $j\ne a+b$ we have
\begin{equation}\label{A-matMup-ww}
\circN_{j,a+b}^{(2,2)}(\buc;\bub;\{\bvc,w\}) = \frac{t(w,\uc_j)h(w,\buc)}{h(w,\bub)}, \qquad j=1,\dots,a,
\end{equation}
\begin{equation}\label{A-matMdown-ww}
\circN_{a+j,a+b}^{(2,2)}(\bub;\{\bvc,w\};\bvb) =0, \qquad j=1,\dots,b-1,
\end{equation}
and sending there $w$ to infinity we obtain that $\circN_{j,a+b}^{(2,2)}\to 0$ as $w\to \infty$ for $j<a+b$.

We see that the last column of the matrix $\circN_{j,k}^{(2,2)}$ contains only one non-zero element
$\circN_{a+b,a+b}^{(2,2)}=-1$. Thus, the determinant in \eqref{F32-22b} reduces to the determinant of the $(a+b-1)\times(a+b-1)$ matrix
with the matrix elements \eqref{A-limmatMup-w} and \eqref{A-limmatMdown-w}. Obviously, this representation coincides with
\eqref{A-UFF-23}, \eqref{A-MatM23}.

{\sl Remark}.
In all considerations above we assumed that the Bethe parameters of on-shell vectors $\mathbb{C}_{a',b'}(\buc;\bvc)$ and
$\mathbb{B}_{a,b}(\bub;\bvb)$ were finite. However, if $r_k(z)\to 1$ at $z\to\infty$, then Bethe equations \eqref{AEigenS-1}
admit infinite solutions as well. The peculiarity of such infinite roots is that the corresponding Bethe vectors are no longer
singular vectors of the zero modes $T_{ij}[0]$ with  $i>j$ (respectively, the operators $T_{ij}[0]$ with  $i<j$ do not
annihilate dual on-shell vectors with infinite parameters). This property played an essential role in our derivations, therefore
one might have impression that the case of infinite Bethe roots  requires a special study. However, as it was shown in
\cite{PakRS15a} for the models with $\mathfrak{gl}(3)$-invariant $R$-matrix, all relations between the form factors remain valid even
in the presence of infinite Bethe parameters. The method of the work \cite{PakRS15a} can be used for the models described by
the $\mathfrak{gl}(2|1)$ superalgebra without any changes. Therefore we do not give here a special consideration to this problem.

\section{Form factors in the models described by $\mathfrak{gl}(1|2)$ superalgebra\label{S-FF-gl12}}

We have mentioned already that the Yangians $Y\bigl(\mathfrak{gl}(1|2)\bigr)$ and $Y\bigl(\mathfrak{gl}(2|1)\bigr)$ are
isomorphic \cite{PakRS16a}. This isomorphism allows us to apply our results to the models with $\mathfrak{gl}(1|2)$ symmetry.

To distinguish between two superalgebras we use the symbol tilde for all the objects related to the $\mathfrak{gl}(1|2)$ superalgebra.
In particular, we use the grading $\widetilde{[1]}=0$ and $\widetilde{[2]}=\widetilde{[3]}=1$. The monodromy matrix entries will be denoted
by $\widetilde{T}_{ij}$, their vacuum eigenvalues by $\widetilde{\lambda}_j$, the Bethe vectors by $\widetilde{\mathbb{B}}_{a,b}(\bu,\bv)$,
and so on.

The isomorphism $\varphi: Y\bigl(\mathfrak{gl}(2|1)\bigr)\to Y\bigl(\mathfrak{gl}(1|2\bigr)$ is defined as follows.
\begin{Def} Let $\bar\jmath=4-j$. Then
\be{def:vph}
\varphi\ :\quad\left\{ \begin{aligned}
\, [j] \quad\ &\to\quad \widetilde{[j]}=[\bar\jmath]+1,\\[1ex]
T_{ij}(u)\quad  &\to\quad  (-1)^{[j][i]+[j]+1}\,\widetilde T_{\bar\jmath,\bar\imath}(u)\\[1ex]
\lambda_j(u)\quad\  &\to\quad \widetilde{\lambda}_{j}(u) =-\lambda_{\bar\jmath}(u).
\end{aligned}\right.
\ee
Hereby,
\be{phiAB}
\varphi(AB)=\varphi(A)\varphi(B).
\ee
\end{Def}

{\sl Remark.} There is a big freedom in the definition of $\varphi$. Namely, we can use the following action
$T_{ij}(u)\to  (-1)^{[j][i]+\alpha[i]+\beta[j]+\gamma}\,\widetilde T_{\bar\jmath,\bar\imath}(u)$, where
$\alpha$, $\beta$, and $\gamma$ are arbitrary constants. Indeed, if the operators $\widetilde T_{ij}(u)$ satisfy
the commutation relations of $Y\bigl(\mathfrak{gl}(1|2)\bigr)$, then  multiplication by $(-1)^{\alpha[i]+\beta[j]+\gamma}$
is equivalent to the multiplication of the monodromy matrix $\widetilde T$ by diagonal twists (from the left by
$\diag((-1)^{\alpha[i]})$ and from the right by $\diag((-1)^{\beta[j]+\gamma})$).
It is clear that after this multiplication the commutation relations are preserved. We have used this possibility
in \eqref{def:vph} in order
to have
\be{phi-trace}
\varphi\bigl(\str(T(u)\bigr)=\str \widetilde T(u).
\ee
However, even this additional restriction does not fix completely the action of $\varphi$. We could choose, for instance,
$T_{ij}(u)\to  (-1)^{[j][i]+[i]+1}\,\widetilde T_{\bar\jmath,\bar\imath}(u)$.

\subsection{Bethe vectors}

Bethe vectors in $Y\bigl(\mathfrak{gl}(1|2)\bigr)$  were constructed in \cite{PakRS16a}:
\be{BV-12}
\widetilde{\mathbb{B}}_{a,b}(\bu;\bv)=(-1)^a\sum
\frac{g(\bu_{\so},\bv_{\so})f(\bv_{\so},\bv_{\st})g(\bu_{\st},\bu_{\so})h(\bv_{\so},\bv_{\so})}
{\widetilde{\lambda}_{2}(\bu_{\st})\widetilde{\lambda}_{2}(\bv)f(\bu,\bv)}
\widetilde{\bT}_{13}(\bv_{\so}) \widetilde{T}_{23}(\bv_{\st}) \widetilde{\bT}_{12}(\bu_{\st}) \widetilde{\Omega}.
\ee
The dual vectors have the following explicit form
\be{dBV-12}
\widetilde{\mathbb{C}}_{a,b}(\bu;\bv)=(-1)^{\frac{a(a-1)}2}\sum
\frac{g(\bu_{\so},\bv_{\so})f(\bv_{\so},\bv_{\st})g(\bu_{\st},\bu_{\so})h(\bv_{\so},\bv_{\so})}
{\widetilde{\lambda}_{2}(\bu_{\st})\widetilde{\lambda}_{2}(\bv)f(\bu,\bv)}
\widetilde{\Omega}^\dagger\widetilde{\bT}_{21}(\bu_{\st})  \widetilde{T}_{32}(\bv_{\st}) \widetilde{\bT}_{31}(\bv_{\so}) .
\ee
Then, assuming that $\varphi(\Omega)=\widetilde{\Omega}$ and $\varphi(\Omega^\dagger)=\widetilde{\Omega}^\dagger$ we find
\be{phiBC}
\varphi\bigl(\mathbb{B}_{a,b}(\bu;\bv)\bigr)= \widetilde{\mathbb{B}}_{b,a}(\bv;\bu),\qquad
\varphi\bigl(\mathbb{C}_{a,b}(\bu;\bv)\bigr)= \widetilde{\mathbb{C}}_{b,a}(\bv;\bu).
\ee
Here we have (dual) Bethe vectors of $Y\bigl(\mathfrak{gl}(2|1)\bigr)$  in the l.h.s., and (dual) Bethe vectors of $Y\bigl(\mathfrak{gl}(1|2)\bigr)$
in the r.h.s. One can also easily check that
\be{psiBC}
\psi\bigl(\widetilde{\mathbb{B}}_{a,b}(\bu;\bv)\bigr)= (-1)^a\widetilde{\mathbb{C}}_{a,b}(\bu;\bv),\qquad
\psi\bigl(\widetilde{\mathbb{C}}_{a,b}(\bu;\bv)\bigr)= \widetilde{\mathbb{B}}_{a,b}(\bu;\bv).
\ee

\subsection{Form factors}

Form factors of the operators $T_{ij}(z)$ depend on the functions $\lambda_k(z)$.
Therefore they are not invariant under the action of $\varphi$:
\be{phi-FF0}
\varphi\Bigl(\mathcal{F}^{(i,j)}\left(z\Bigl|\begin{smallmatrix}
\buc & \bub \\ \bvc & \bvb \end{smallmatrix}\right)^{a',a}_{b',b}\Bigr)
=\mathcal{F}^{(i,j)}\left(z\Bigl|\begin{smallmatrix}
\buc & \bub \\ \bvc & \bvb \end{smallmatrix}\right)^{a',a}_{b',b}\Bigr|_{\lambda_k(z)\rightarrow -\lambda_{\bar k}(z)}.
\ee
On the other hand we have
\begin{multline}\label{phi-FF1}
\varphi\left(\mathcal{F}^{(i,j)}\left(z\Bigl|\begin{smallmatrix}
\buc & \bub \\ \bvc & \bvb \end{smallmatrix}\right)^{a',a}_{b',b}\right)=
\varphi\Bigl(\mathbb{C}_{a',b'}(\buc;\bvc)T_{ij}(z) \mathbb{B}_{a,b}(\bub;\bvb)
\Bigr)\\
=(-1)^{[j][i]+[j]+1 }
\widetilde{\mathbb{C}}_{b',a'}(\bvc;\buc)\widetilde{T}_{\bar\jmath,\bar\imath}(z)\widetilde{\mathbb{B}}_{b,a}(\bvb;\bub)
=(-1)^{[j][i]+[j]+1 }
\widetilde{\mathcal{F}}^{(\bar\jmath,\bar\imath)}\left(z\Bigl|\begin{smallmatrix}
\bvc & \bvb \\ \buc & \bub \end{smallmatrix}\right)^{b',b}_{a',a}.
\end{multline}
Thus, we obtain
\be{ff-ff1}
(-1)^{[j][i]+[j]+1 }\widetilde{\mathcal{F}}^{(\bar\jmath,\bar\imath)}\left(z\Bigl|\begin{smallmatrix}
\bvc & \bvb \\ \buc & \bub \end{smallmatrix}\right)^{b',b}_{a',a}=
\mathcal{F}^{(i,j)}\left(z\Bigl|\begin{smallmatrix}
\buc & \bub \\ \bvc & \bvb \end{smallmatrix}\right)^{a',a}_{b',b}\Bigr|_{\lambda_k(z)= -\widetilde\lambda_{\bar k}(z)}.
\ee
Changing here
\be{chch}
\bucb\leftrightarrow\bvcb,\qquad a \leftrightarrow b,\qquad a' \leftrightarrow b',\qquad \bar\jmath \leftrightarrow i,\qquad
\bar\imath \leftrightarrow j,
\ee
 we find
\be{FF21-120}
\widetilde{\mathcal{F}}^{(i,j)}\left(z\Bigl|\begin{smallmatrix}
\buc & \bub \\ \bvc & \bvb \end{smallmatrix}\right)^{a',a}_{b',b}
=(-1)^{[\bar\jmath][\bar\imath]+[\bar\imath]+1}\mathcal{F}^{(\bar\jmath,\bar\imath)}\left(z\Bigl|\begin{smallmatrix}
\bvc & \bvb \\ \buc & \bub \end{smallmatrix}\right)^{b',b}_{a',a}\Bigr|_{\lambda_k(z)= -\widetilde\lambda_{\bar k}(z)}.
\ee
It remains to use $\widetilde{[j]}=[\bar\jmath]+1$, and we finally arrive at
\be{FF21-12}
\widetilde{\mathcal{F}}^{(i,j)}\left(z\Bigl|\begin{smallmatrix}
\buc & \bub \\ \bvc & \bvb \end{smallmatrix}\right)^{a',a}_{b',b}
=(-1)^{\widetilde{[j]}\widetilde{[i]}+\widetilde{[j]}+1}\mathcal{F}^{(\bar\jmath,\bar\imath)}\left(z\Bigl|\begin{smallmatrix}
\bvc & \bvb \\ \buc & \bub \end{smallmatrix}\right)^{b',b}_{a',a}\Bigr|_{\lambda_k(z)= -\widetilde\lambda_{\bar k}(z)}.
\ee
Thus, the form factors of the monodromy matrix entries in the models with $\mathfrak{gl}(1|2)$ and $\mathfrak{gl}(2|1)$ symmetries are
related to each other by the replacement of variables \eqref{chch}.

\section*{Conclusion}

In this paper we obtained determinant representations for form factors of the monodromy matrix entries
in integrable models described by $\mathfrak{gl}(2|1)$ and $\mathfrak{gl}(1|2)$ superalgebras.
The method is based on the determinant formula for a particular case of Bethe vectors scalar product \cite{HutLPRS16c}. This formula allows one
to calculate form factors of the diagonal operators $T_{ii}$. Further calculation of form factors of the off-diagonal operators $T_{ik}$ is
based on the zero modes method \cite{PakRS15a}.

The obtained results can be used for the calculation of form factors and correlation functions in the supersymmetric t-J model.
For this model the solution of the quantum inverse scattering problem is known \cite{MaiT00,GohK00}. Therefore, form
factors of local operators  can be easily reduced to the ones considered in
the present paper.

The calculation of form factors in models with $\mathfrak{gl}(m|n)$ symmetry remains to be done. Any results in this
field would be desirable in view of their possible application to Hubbard model and supersymmetric gauge theories.   It is clear that
the zero modes method works in this case as well. Therefore, it would be enough to obtain a determinant formula for
only one form factor. All other form factors would be achieved form the initial one as special limits of the Bethe parameters. However,
the problem of calculating the initial form factor meets  serious technical difficulties.

\section*{Acknowledgements}
N.A.S. thanks LAPTH in Annecy-le-Vieux for the hospitality and stimulating scientific atmosphere, and CNRS for partial financial support.
The work of A.L.  has been funded by the  Russian Academic Excellence Project 5-100 and by joint NASU-CNRS project F14-2016.
The work of S.P. was supported in part by the RFBR grant 16-01-00562-a.
N.A.S. was  supported by  the grants RFBR-15-31-20484-mol-a-ved and RFBR-14-01-00860-a.

\end{document}